\documentclass[11pt]{article}
\usepackage{acl}

\usepackage{color}
\usepackage{bbm}
\usepackage{multirow}
\usepackage[inline]{enumitem}
\usepackage{sistyle}
\usepackage[ruled]{algorithm2e}
\usepackage{booktabs}
\usepackage{caption}
\SIthousandsep{,}
\usepackage{makecell}
\usepackage[skip=0pt]{caption}
\usepackage{hyperref}
\usepackage{array} 
\usepackage{graphicx}
\usepackage[table]{xcolor} 
\usepackage{xcolor}
\usepackage{acronym}
\usepackage[export]{adjustbox}
\usepackage{subcaption}
\usepackage{arydshln} 
\usepackage{booktabs}
\usepackage{multirow}
\usepackage{geometry}
\usepackage{amsmath,amssymb}
\usepackage{times}
\usepackage{latexsym}

\title{Can LLM Rerankers Predict Their Own Ranking Performance?}
\author{
Shiyu Ni\textsuperscript{\rm 1,2,3} \quad
\textbf{Keping Bi}\textsuperscript{\rm 1,2,3} \\
\textbf{Jiafeng Guo}\textsuperscript{\rm 1,2,3} \quad
\textbf{Jingtong Wu} \quad
\textbf{Zengxin Han} \quad
\textbf{Xueqi Cheng}\textsuperscript{\rm 1,2,3}\\
\textsuperscript{\rm 1} State Key Laboratory of AI Safety \\
\textsuperscript{\rm 2} Institute of Computing Technology, Chinese Academy of Sciences \\
\textsuperscript{\rm 3} University of Chinese Academy of Sciences \\
}

\begin{document}
\maketitle



\begin{abstract}
Retrieval effectiveness varies substantially across queries, making it important to estimate ranking quality before relevance judgments are available. Query performance prediction (QPP) addresses this need, but most existing methods rely on external predictors after retrieval or reranking. In this paper, we study \textit{reranker-internal QPP}: can an LLM reranker estimate the quality of the ranking it has just produced? We investigate both training-free and training-based approaches. For training-free estimation, we examine metric-specific self-consistency across sampled rankings and verbalized confidence produced directly by the reranker. Experiments on TREC Deep Learning 2019--2022 with four LLMs show that self-consistency is competitive with the state-of-the-art (SOTA) approach and better calibrated in almost all settings, while direct verbalized confidence is severely overconfident. To improve verbalized confidence, we propose two supervised methods, Verb-Num and Verb-List, which enable LLM rerankers to produce calibrated ranking-quality estimates with only a few additional output tokens.\looseness=-1
\end{abstract}


\begin{figure}
    \centering
    \includegraphics[width=0.98\linewidth]{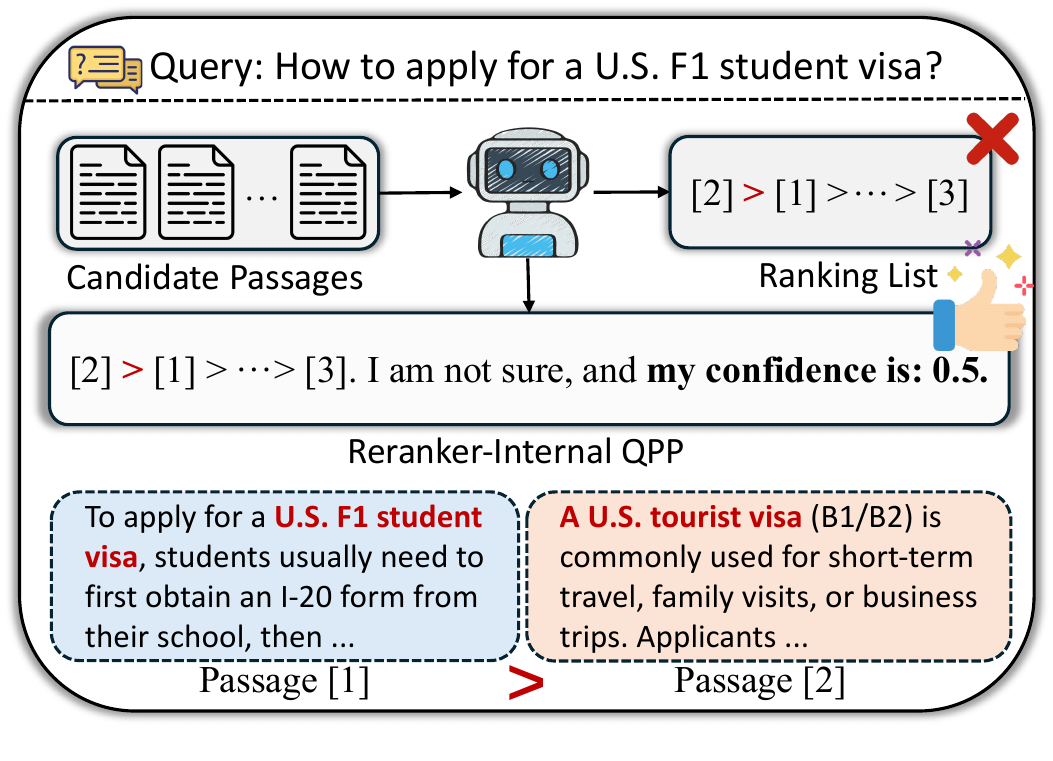}
    \caption{Overview of reranker-internal QPP. In this example, the model incorrectly ranks the less relevant Passage 2 ahead of Passage 1. We expect the reranker not only generates the ranked list but also expresses its confidence, helping prevent users from being misled by an erroneous ranking result.}
    \label{fig:s2s_ranking}
\end{figure}

\section{Introduction}
Retrieval effectiveness varies substantially across queries. Some queries are specific and well covered by the collection, while others are ambiguous, underspecified, affected by vocabulary mismatch, or have few relevant documents. A system that performs well on average can therefore still return poor results for particular queries, and these failures are especially consequential when top-ranked passages are shown to users or passed to retrieval-augmented LLMs. Query performance prediction (QPP) addresses this problem by estimating, before or without relevance judgments, how well a retrieval system is likely to perform for a given query~\citep{carmel2010estimating}. Such estimates make retrieval more reliable and adaptive: they can identify difficult queries, trigger query reformulation or stronger reranking only when needed, allocate computation to uncertain cases, and support monitoring when manual relevance judgments are unavailable.

Existing QPP methods are commonly grouped into pre-retrieval and post-retrieval approaches~\citep{carmel2010estimating}, with the latter usually stronger because they exploit the returned list. Unsupervised methods rely on signals such as query clarity~\citep{cronen2002predicting}, score distributions~\citep{shtok2012predicting,perez2010standard,tao2014query}, robustness, document coherence~\citep{aslam2007query,arabzadeh2021query}, and, for neural rankers, score uncertainty or representation-based agreement~\citep{zendel2023entropy,vlachou2023coherence,singh2023unsupervised}. Supervised methods learn from labeled queries using neural or PLM-based representations of queries and ranked documents~\citep{hashemi2019performance,arabzadeh2021bert,datta2022pointwise}. QPP-Gen~\citep{meng2025query} recently introduced LLM-based relevance judgments for QPP. Most of these methods, however, remain separate modules placed after retrieval or reranking. \looseness=-1

This paper studies a different question: can an LLM reranker estimate the quality of the ranking it has just produced? We refer to this setting as \textbf{reranker-internal QPP}, emphasizing that the confidence signal comes from the reranker itself rather than from an external QPP model. The question is natural for sequence-to-sequence LLM rerankers, which take a query and candidate passages as input and directly generate a ranked list (Figure~\ref{fig:s2s_ranking}). If such a model can judge document relevance well enough to rank, it may also contain useful signals about when its ranking is likely to be reliable. \looseness=-1

Reranker-internal QPP is closely related to confidence calibration in question answering (QA), where a calibrated model should assign confidence that matches answer correctness. Prior work has studied probability-based confidence~\citep{guo2017calibration,jiang2021can,kadavath2022language,si2022prompting}, self-consistency~\citep{Manakul2023SelfCheckGPT,kuhn2023semantic}, and verbalized confidence~\citep{tian2023just,xiong2023can,ni2024llms}, often finding that LLMs are overconfident. QPP shares the same calibration goal, but ranking differs from QA: it evaluates an ordered list rather than a single answer, and quality is position-sensitive and metric-dependent, e.g., Precision, MRR, or NDCG. Thus, answer-level agreement does not directly transfer to rankings, and verbalized confidence must estimate performance over multiple ordered documents.

We investigate both training-free and training-based approaches. For training-free estimation, we study \textit{self-consistency} across sampled rankings with metric-specific measures for Precision, MRR, and NDCG, and \textit{verbalized confidence}, where the model directly outputs a numerical confidence score. We then study whether supervised training can teach an LLM reranker to express calibrated confidence while preserving efficient reranking. \looseness=-1

\noindent\textbf{Research Questions.}
We organize the study around three questions:
\begin{itemize}[leftmargin=*,itemsep=0pt,topsep=2pt,parsep=0pt]
\item \textbf{RQ1}: How predictive and calibrated are metric-specific self-consistency signals for LLM reranking performance?
\item \textbf{RQ2}: Can LLM rerankers verbalize accurate confidence in their own ranking quality without training, and what failure modes arise?
\item \textbf{RQ3}: Can supervised training enable LLM rerankers to output calibrated ranking-quality estimates?
\end{itemize}

\noindent\textbf{Main Findings.}
Experiments on TREC DL 2019--2022 with LLaMA3.1 and Qwen2.5 models show that self-consistency is a strong training-free signal: it achieves competitive Spearman correlation with QPP-Gen and is better calibrated in almost all settings, though it can underperform on precision-oriented metrics. In contrast, direct verbalized confidence is poorly calibrated, with LLM rerankers assigning high confidence to most queries. We further propose two supervised methods: \textbf{Verb-Num}, which outputs a scalar estimate of a target ranking metric, and \textbf{Verb-List}, which outputs relevance indicators for top-ranked documents. Both improve QPP performance; Verb-Num tends to obtain stronger discriminability, while Verb-List is better calibrated and requires only a few additional output tokens.

\noindent\textbf{Contributions.}
Our contributions are:
1) We identify the connection between QPP and confidence calibration in QA, and formulate reranker-internal QPP for sequence-to-sequence LLM rerankers.
2) We analyze two training-free confidence signals, metric-specific self-consistency and verbalized confidence, showing that the former is competitive and better calibrated than QPP-Gen, while the latter suffers from severe overconfidence.
3) We propose Verb-Num and Verb-List, two training methods that enable LLM rerankers to produce calibrated ranking-quality estimates with minimal additional generation cost.

\section{Related Work}
\paragraph{Query Performance Prediction}
Post-retrieval QPP estimates retrieval effectiveness without relevance labels. 
Existing methods are commonly divided into unsupervised and supervised approaches. 
Unsupervised methods exploit properties of retrieved results, such as clarity~\citep{cronen2002predicting}, robustness~\citep{aslam2007query}, coherence~\citep{arabzadeh2021query,vlachou2023coherence}, ranking scores~\citep{shtok2012predicting}, entropy~\citep{zendel2023entropy}, and pairwise preferences~\citep{singh2023unsupervised}. 
Supervised methods learn performance estimators from ranked lists, ranging from neural models~\citep{zamani2018neural,datta2022deep} to fine-tuned PLMs and multi-task frameworks~\citep{arabzadeh2021bert,chen2022groupwise,khodabakhsh2023learning}. 
Recently, QPP-Gen~\citep{meng2025query} leverages LLMs to judge document relevance and achieves strong performance. 
Unlike prior methods that treat ranking and QPP as separate modules, we study whether an LLM reranker can perceive its own ranking performance. \looseness=-1

\paragraph{LLM Knowledge Boundary Perception}
A reliable model should recognize when its outputs are likely to be wrong. 
This ability is often studied through confidence estimation, including probabilistic confidence from token likelihoods~\citep{guo2017calibration,kadavath2022language}, self-consistency across generations~\citep{Manakul2023SelfCheckGPT,kuhn2023semantic}, verbalized confidence~\citep{xiong2023can,tian2023just}, and confidence signals from internal representations~\citep{azaria2023internal,chen2024inside,ni2025towards}. 
Our work extends this perspective to LLM reranking, asking whether the reranker can assess the quality of the ranking it just produced.
Due to space limitations, we provide a more detailed discussion in~\S~\ref{app:related_work}. \looseness=-1


\begin{figure*}[h!]
    \centering
    \includegraphics[width=1.0\linewidth]{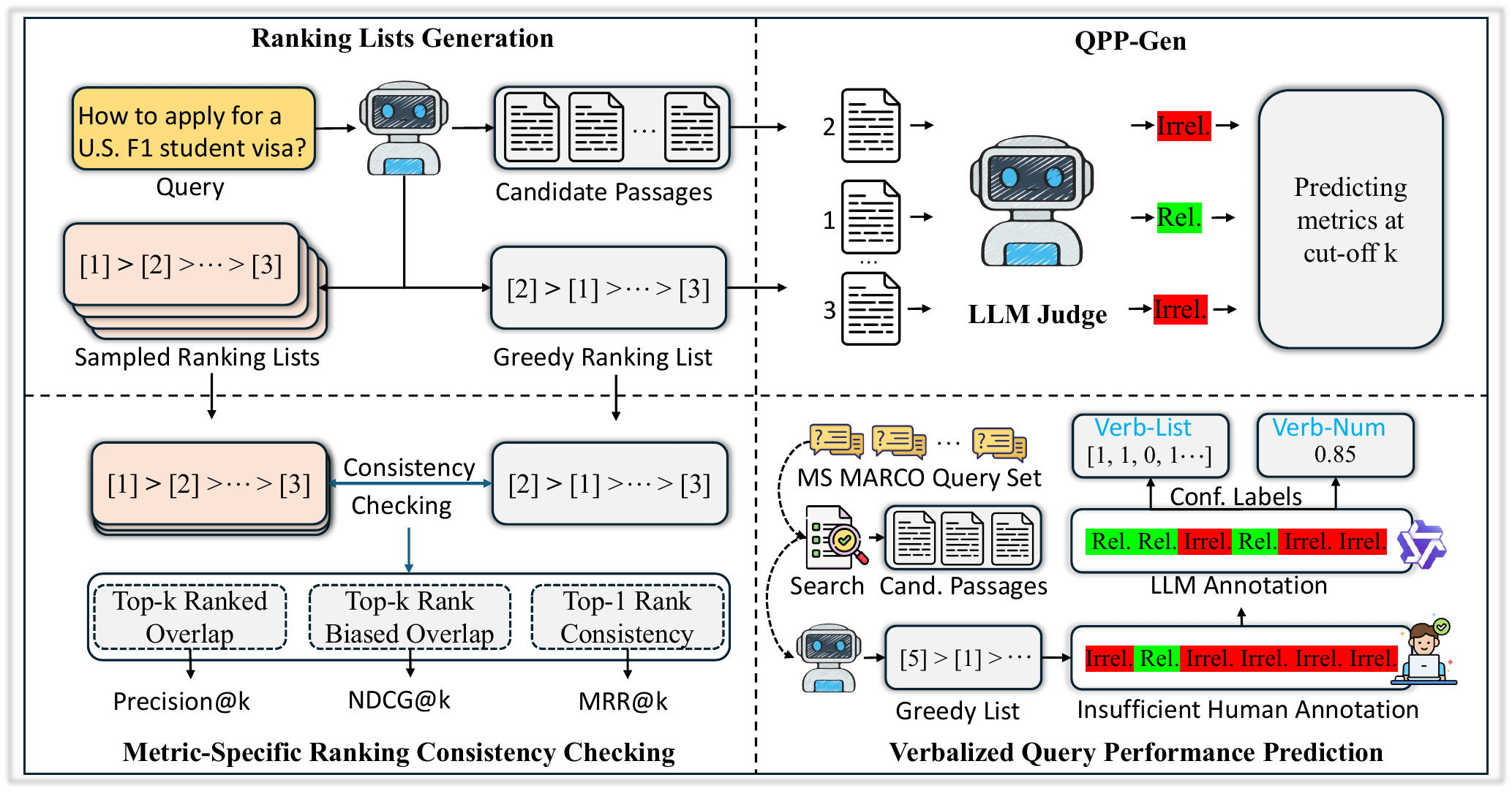}
    \caption{Overview of ranking lists generation, QPP-Gen, metric-specific consistency checking, and verbalized ranking-quality estimates calibration. QPP-Gen uses an LLM for pointwise relevance judgments. Self-consistency-based confidence measures agreement between multiple ranking lists under different IR metrics. Verbalized training labels are obtained by assessing rankings with both human annotation and LLM annotation.}
    \label{fig:method}
\end{figure*}

\section{Preliminary}
\label{sec:preliminary}
\noindent{\textbf{Sequence-to-sequence Reranking.}} Following~\cite{liu2025sliding}, for a query $q$, we first retrieve the top 100 passages $\mathcal{D}=\{d_1, d_2, \dots, d_{100}\}$ using BM25~\cite{robertson1995okapi}. An LLM reranker then takes all passages as input and outputs a relevance-based ordering in text, as shown in Figure~\ref{fig:prompt_ranking}. We represent the resulting ranking as a sequence of indices $\mathcal{I} = \{i_1, i_2, \dots, i_{100}\}$. For example, $i_1=5$ indicates that the 5th passage in the original input is ranked first after re-ranking.

\noindent{\textbf{QPP-Gen.}} Query performance prediction aims to assess the quality of a ranked list without ground-truth relevance annotations. QPP-Gen achieves SOTA performance among existing unsupervised QPP methods. It leverages an LLM to perform relevance judgments on each passage in the ranking list and calculates the corresponding ranking performance based on these judgments, as showing in Figure~\ref{fig:method}. \looseness=-1

\section{Experimental Setup}
\noindent{\textbf{Ranking Quality Evaluation.}}  
We evaluate ranking quality with standard top-$k$ IR metrics, including \textbf{Precision@$k$}, \textbf{MRR@$k$} (Mean Reciprocal Rank), and \textbf{NDCG@$k$}. Since sequence-to-sequence reranking only reorders the retrieved candidate pool $\mathcal{D}$, we compute NDCG with respect to the relevant passages within this pool and denote it as $\textbf{NDCG-I}$ where ``I" denotes ``Input".
Let $r_j \in \{0,1\}$ be the binary relevance label of passage $d_j$, and let $\mathcal{R}=\{j:r_j=1\}$ be the set of relevant passage indices in $\mathcal{D}$. Given the generated ranking $\mathcal{I}=\{i_1,\dots,i_{100}\}$, the metrics are:
\begin{equation*}
\text{NDCG-I}@k = \frac{\text{DCG@}k}{\text{IDCG-I}@k},
\end{equation*}
\begin{equation*}
\text{DCG@}k = \sum_{m=1}^{k} \frac{\mathbb{I}(i_m \in \mathcal{R})}{\log_2(m+1)},
\end{equation*}
\begin{equation*}
\text{IDCG-I}@k = \sum_{m=1}^{ k} \frac{1}{\log_2(m+1)} .
\end{equation*}
Here, $i_m$ denotes the passage index ranked at position $m$, and $\mathbb{I}(\cdot)$ is the indicator function.

\noindent\textbf{QPP Evaluation}
Unlike prior QPP studies~\cite{meng2025query,arabzadeh2021bert}, which primarily focus on discriminability, we argue that a good QPP method should also be well calibrated, i.e., its predicted scores should numerically reflect the actual query performance. Therefore, in addition to evaluating discriminability with the \textbf{Spearman correlation coefficient} ($\rho$)~\cite{spearman1961proof}, we introduce the \textbf{Expected Calibration Error (ECE)}~\cite{guo2017calibration} to explicitly assess calibration quality.
ECE measures the discrepancy between predicted scores and true performance values. Specifically, samples are partitioned into $B$ bins according to their predicted scores, and the calibration error is computed as the weighted average absolute difference between the average predicted score and the average true performance within each bin (B=10 in the paper):
\begin{equation}
   \text{ECE} = \sum_{b=1}^{B} \frac{|S_b|}{S} \left| \text{real}(b) - \text{pred}(b) \right|,
\end{equation}
where $S_b$ denotes samples in bin $b$, and $\text{real}(b)$ and $\text{pred}(b)$ denote the average true and predicted scores in that bin, respectively. Lower ECE indicates better calibration. \looseness=-1

\noindent{\textbf{Datasets}.}
Following the previous study~\cite{meng2025query}, we conduct experiments on TREC Deep Learning (TREC-DL) tracks from 2019 to 2022~\cite{craswell2025overview}, which are standard benchmarks for passage ranking. We also binarize relevance labels by treating passages with relevance $\ge$ 2 as positive.  Details can be seen in~\S~\ref{app:datasets}.  \looseness=-1


\noindent{\textbf{Models}.} We use two representative series of LLMs, namely LLaMA3.1~\cite{grattafiori2024llama} and Qwen2.5~\cite{team2024qwen2}. For LLaMA3.1, we use LLaMA3.1-8B-Instruct. To study the effect of model size, we leverage Qwen2.5-Instruct ranging from 7B to 32B.

\section{QPP via Ranking Consistency}

\label{sec:self-consistency}
In this section, we answer how predictive and calibrated are metric-specific self-consistency signals for LLM reranking performance (\textbf{RQ1}).

\subsection{Consistency between Two Ranking Lists}

Measuring self-consistency between two ranking lists is non-trivial. First, ranking quality can be evaluated from multiple perspectives, a single consistency measure can not fully capture these distinct aspects. Second, it needs to account for positional differences.  
To address these challenges, we leverage the characteristics of established IR metrics to design metric-driven consistency measures. \looseness=-1
\looseness=-1

\noindent{\textbf{Ranking Lists Generation.}} The model is asked to generate two types of ranking lists:
\begin{itemize}[leftmargin=*]
    \item \textbf{Greedy Ranking (\(\mathcal{I}_g\))}: Generated using greedy decoding (temperature $T=0$). This is used for ranking performance evaluation.
    \item \textbf{Sampled Rankings (\(\mathcal{I}_{s,j}\))}: Generating $N=20$ ranking lists by setting $T=1$, where $s$ denotes ``sampling" and $j \in \{1, \dots, N\}$.
\end{itemize}

\noindent{\textbf{Metric-Specific Ranking Consistency Checking.}}
For each  IR metric, we design a specific consistency function \(\text{Consis}(\mathcal{I}_1, \mathcal{I}_2)\) to quantify the similarity between two ranking lists \(\mathcal{I}_1\) and \(\mathcal{I}_2\).

\begin{itemize}[leftmargin=*]
    \item \textbf{Consistency for Precision@k}: 
    Since Precision considers the top-$k$ results as an unordered set, we measure its consistency by the overlap ratio of the top-$k$ items:
    \begin{equation*}
    \text{Consis}_{\text{Prec}}(\mathcal{I}_1, \mathcal{I}_2) = \frac{|\mathcal{I}_1^{1:k} \cap \mathcal{I}_2^{1:k}|}{k}.
    \end{equation*}

    \item \textbf{Consistency for \(\text{NDCG-I@k}\)}: 
    To align with the position-based discounting, we adopt Rank Biased Overlap (RBO)~\cite{webber2010similarity}. RBO evaluates the similarity of two lists by calculating the \textit{agreement} \(A_d\) at each depth \(d\), which represents the proportion of shared items within the top-\(d\) items:
    \begin{equation*}
    A_d = \frac{|\mathcal{I}_{1,1:d} \cap \mathcal{I}_{2,1:d}|}{d},
    \end{equation*}
    where \(\mathcal{I}_{1,1:d}\) and \(\mathcal{I}_{2,1:d}\) are the sets of the first \(d\) documents in each list.
    As we evaluate ranking performance at a fixed depth \(k\), we employ a truncated version of RBO and normalize the score by the total weight of the first \(k\) terms:
    \begin{equation*}
    \text{Consis}_{\text{NDCG-I@k}}(\mathcal{I}_1, \mathcal{I}_2) = \frac{1-p}{1-p^k} \sum_{d=1}^{k} p^{d-1} \cdot A_d,
    \end{equation*}
    where \(p \in (0, 1)\) is the parameter (i.e., \(p=0.9\) in this paper) that determines the weight concentration on higher-ranked items; a smaller \(p\) assigns more weight to items at earlier positions and \({1-p^k}\) is used for normalization.

    \item \textbf{Consistency for MRR@k}: 
    Mirroring MRR's sensitivity to the position of the first relevant document, we view the first-ranked item in (\(\mathcal{I}_1\)) as relevant and define consistency as the reciprocal rank of the top-ranked item of the reference list (\(\mathcal{I}_1\)) within the compared list (\(\mathcal{I}_2\)):
    \begin{equation}
    \text{Consis}_{\text{MRR@k}}(\mathcal{I}_1, \mathcal{I}_2) = \frac{1}{\text{Rank}(i_{1,1}, \mathcal{I}_2)}
    \end{equation}
    where \(i_{1,1}\) is the item ranked first in \(\mathcal{I}_1\), and \(\text{Rank}(i_{1,1}, \mathcal{I}_2)\) denotes its rank index in \(\mathcal{I}_2\).
\end{itemize}

\noindent{\textbf{Final Consistency Score.}}
Consistency on a query is computed by the average consistency between the stable greedy ranking \(\mathcal{I}_g\) and each of the \(N\) sampled rankings \(\mathcal{I}_{s,j}\). 
We expect that a model with a clear understanding of ranking should achieve both high self-consistency and strong ranking performance.
\begin{table*}[t]
\centering
\caption{Average QPP performance over TREC DL19--22. Bolds denote the best scores.}
\label{tab:avg_results_consis}
\small
\scalebox{0.9}{
\begin{tabular}{llcccccc}
\toprule
& & \multicolumn{2}{c}{NDCG-I@10} 
& \multicolumn{2}{c}{Precision@10} 
& \multicolumn{2}{c}{MRR@10} \\
\cmidrule(lr){3-4}
\cmidrule(lr){5-6}
\cmidrule(lr){7-8}

Model & Method 
& Spearman $\uparrow$ & ECE $\downarrow$
& Spearman $\uparrow$ & ECE $\downarrow$
& Spearman $\uparrow$ & ECE $\downarrow$ \\
\midrule

\multirow{2}{*}{Llama3.1-8B}
& QPP-Gen
& 0.268 & 0.389
& \textbf{0.436} & 0.496
& 0.180 & 0.282 \\

& Self-Consis
& \textbf{0.337} & \textbf{0.179}
& 0.379 & \textbf{0.166}
& \textbf{0.225} & \textbf{0.221} \\
\midrule

\multirow{2}{*}{Qwen2.5-7B}
& QPP-Gen
& 0.289 & 0.200
& 0.403 & \textbf{0.154}
& \textbf{0.260} & 0.261 \\

& Self-Consis
& \textbf{0.336} & \textbf{0.171}
& \textbf{0.424} & 0.166
& 0.204 & \textbf{0.219} \\
\midrule

\multirow{2}{*}{Qwen2.5-14B}
& QPP-Gen
& 0.371 & 0.208
& \textbf{0.548} & 0.249
& -- & 0.215 \\

& Self-Consis
& \textbf{0.373} & \textbf{0.158}
& 0.423 & \textbf{0.158}
& \textbf{0.307} & \textbf{0.202} \\
\midrule

\multirow{2}{*}{Qwen2.5-32B}
& QPP-Gen
& 0.295 & 0.212
& \textbf{0.544} & 0.251
& \textbf{0.260} & 0.229 \\

& Self-Consis
& \textbf{0.394} & \textbf{0.150}
& 0.454 & \textbf{0.171}
& 0.210 & \textbf{0.197} \\

\bottomrule
\end{tabular}
}
\end{table*}

\begin{figure*}[h!]
    \centering
    \includegraphics[width=\linewidth]{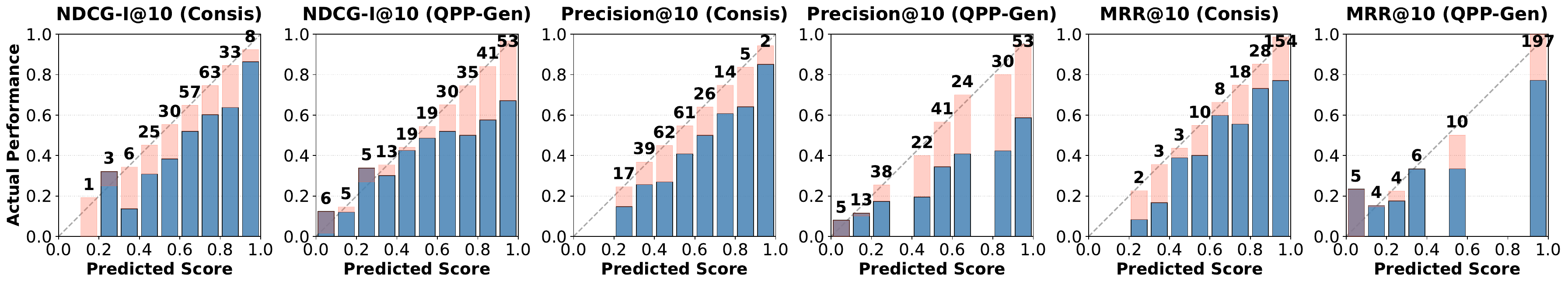}
    \caption{ECE of self-consistency and QPP-Gen on Qwen2.5-14B-Instruct. In each plot, the x-axis represents bins of predicted scores, while the y-axis shows the true performance of samples within each bin. The diagonal line $y = x$ indicates perfect calibration. The numbers above the bars denote the count of samples whose predicted confidence falls into each bin. Results are aggregated over DL19–22.}
    \label{fig:ECE-Qwen14B}
\end{figure*}

\subsection{Baselines}

\looseness=-1
We compare self-consistency with QPP-Gen~\citep{meng2025query}, a SOTA unsupervised QPP baseline, illustrated in Section \ref{sec:preliminary}. 
\noindent\textbf{Why not include traditional unsupervised QPP methods?}
We focus on QPP-Gen as a strong and representative LLM-based method. Traditional score- or embedding-based QPP methods are less applicable to sequence-to-sequence reranking, where LLMs output only ranked lists without calibrated scores or query--document embeddings. Robustness-based QPP could be adapted by perturbing queries and measuring ranking stability, but this would be a separate consistency-based variant and is left for future work. 


\subsection{Results and Analysis}

The ranking performance of each model and the self-consistency scores are shown in Table~\ref{tab:ranking performance}. The performance of self-consistency and QPP-Gen is shown in Table~\ref{tab:avg_results_consis}. Results on each dataset can be seen in Table~\ref{tab:spearman_and_ece}. We find:

\textbf{1) Self-consistency is better calibrated than QPP-Gen.}
As shown in Table~\ref{tab:avg_results_consis}, compared with QPP-Gen, the ECE corresponding to self-consistency is lower in almost all cases. To further understand this phenomenon, we analyze the distributions of the confidence scores predicted by the two methods, as shown in Figure~\ref{fig:ECE-Qwen14B}. It can be observed that the confidence predicted by both methods is generally higher than the actual ranking performance. In particular, QPP-Gen assigns high confidence to more samples, indicating that this method incorrectly judges many highly ranked but irrelevant documents as relevant, which explains its worse ECE.

\textbf{2) Self-consistency achieves competitive correlation with ranking performance.}  
As shown in Table~\ref{tab:avg_results_consis}, compared with QPP-Gen, Self-consis achieves competitive Spearman correlations in most cases. Notably, Self-consis performs better on \(\text{NDCG-I}@10\), while it can underperform on Precision and MRR. This contrast stems from the different limitations of the two methods.

\textit{QPP-Gen is not well-suited for recall-aware metrics.}  
Since top-ranked passages are more likely to be relevant, false-positive errors of QPP-Gen are less emphasized at higher ranks. Consequently, QPP-Gen is naturally well suited for precision-oriented metrics such as Precision@k and MRR. However, recall-aware metrics such as $\text{NDCG-I}@10$ also depend on judgments of lower-ranked passages, where false positives tend to accumulate and reduce predictive accuracy. As a result, QPP-Gen performs worse than self-consistency on $\text{NDCG-I}@10$.

\begin{figure}[ht]
  \centering
  \begin{subfigure}{0.49\linewidth}
    \centering
    \includegraphics[width=\linewidth]{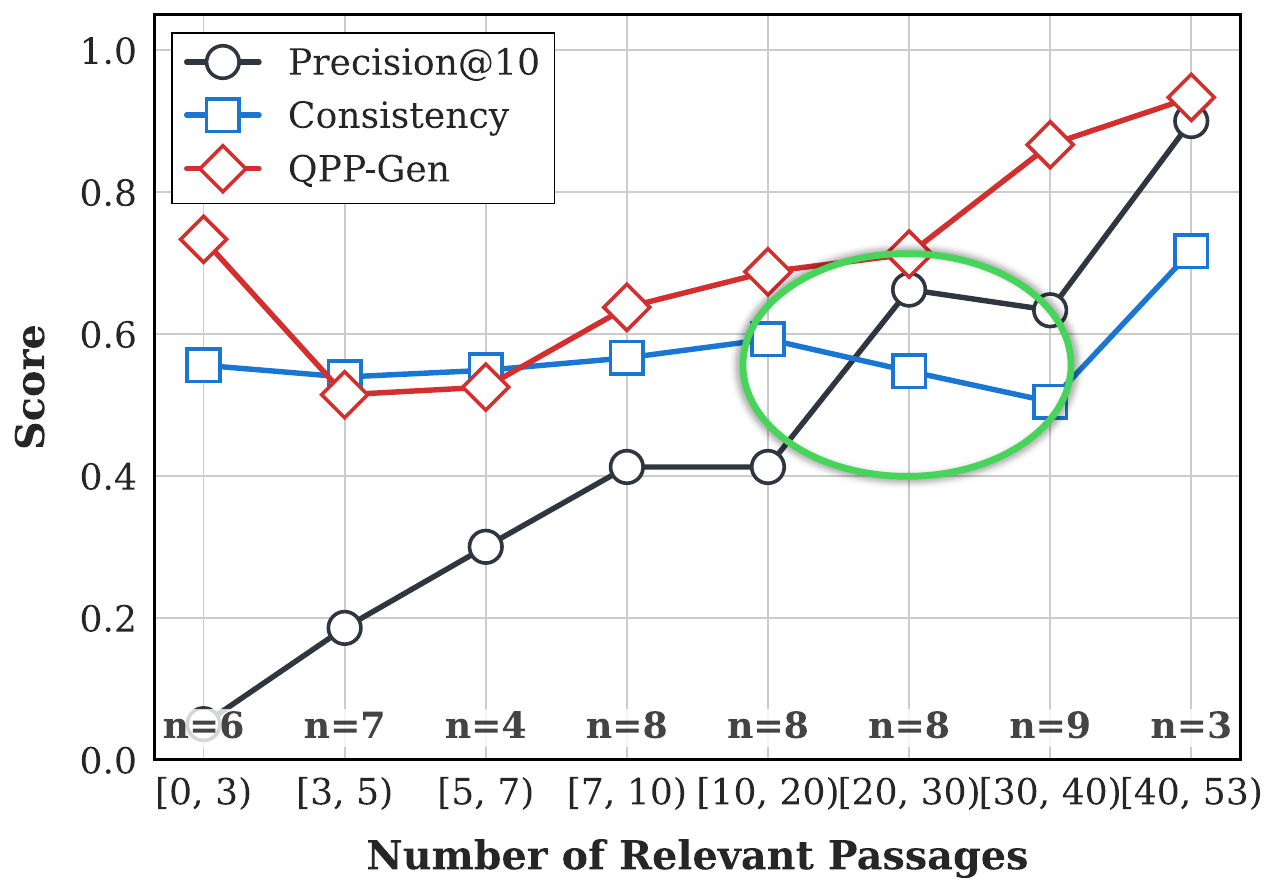}
    \caption{Qwen2.5-14B}
    \label{fig:qwen14b}
  \end{subfigure}
  \hfill
  \begin{subfigure}{0.49\linewidth}
    \centering
    \includegraphics[width=\linewidth]{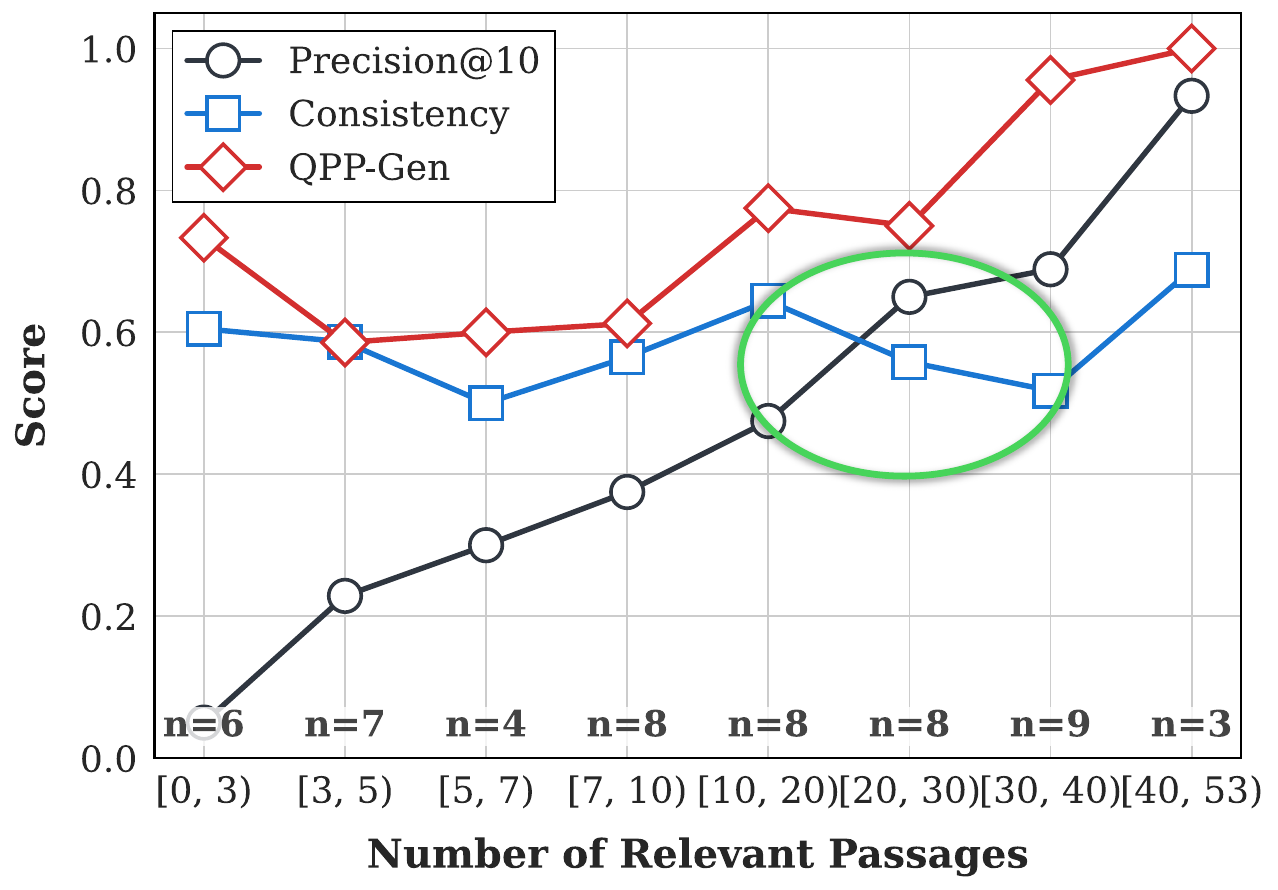}
    \caption{Qwen2.5-32B}
    \label{fig:qwen32b}
  \end{subfigure}
  
  \caption{Trends of self-consistency and QPP-Gen predicted scores for Qwen2.5-14B and 32B on DL21 with respect to the number of relevant passages in the input. }
  \label{fig:conf_with_count_of_relevant_passages}
  \vspace{-0.1cm}
\end{figure}

\textit{Self-consistency has a structural mismatch with ranking quality in relevance-dense pools.}
When many relevant documents are available, the model may place different relevant subsets at the top across runs, yielding high ranking quality but low consistency. 
As a result, self-consistency can underperform QPP-Gen on precision-oriented metrics. 
Figure~\ref{fig:conf_with_count_of_relevant_passages} supports this observation: as the number of relevant documents increases, Precision@k tends to improve, whereas self-consistency may decline.
So stability is not a guarantee of correctness for each query. 
A model can repeatedly make the same relevance mistake while remaining highly consistent. 
Self-consistency also requires multiple ranking runs, making it less efficient than QPP-Gen. \looseness=-1
%

\vspace{-0.3cm}
\begin{figure*}[ht]
    \centering
    \includegraphics[width=\linewidth]{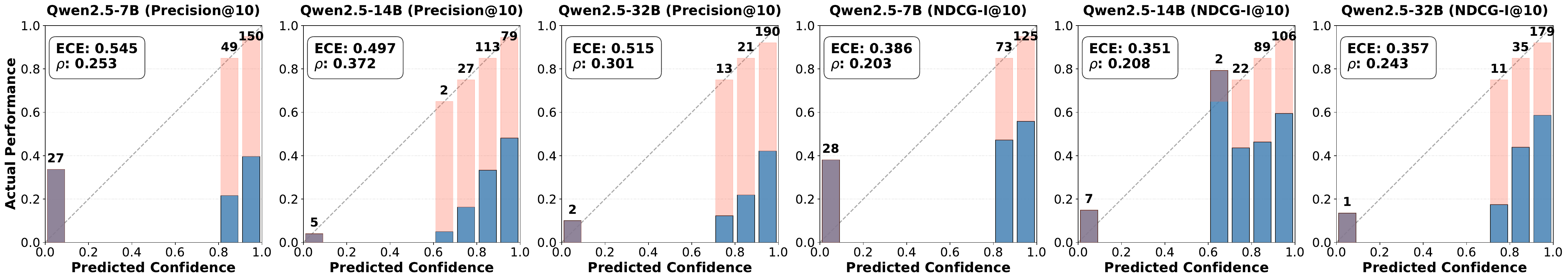}
    \caption{Comparison between verbalized confidence and true performance for the Qwen2.5 series models on the TREC DL19-22 datasets. $\rho$ represents the Spearman correlation coefficient.}
    \label{fig:ECE-Verb}
\end{figure*}
\section{QPP via Verbalized Confidence}
In this section, we answer \textbf{RQ2}: Can LLM rerankers verbalize accurate confidence in their own ranking quality without training, and what failure modes arise?
\vspace{-0.2cm}
\subsection{Experimental Setup}
 The model is asked to output its confidence that reflects a given metric such as Precision@10 while ranking. The output format is: \texttt{[2] $>$ [1] $>$ $\cdots >$ [5]. Confidence: 0.85.} We study Qwen2.5-series models. The other settings are the same as those in ~\S~\ref{sec:self-consistency}.

\subsection{Results and Analysis}
Since the model is required to perform both ranking and confidence estimation, we first investigate whether this dual-task setup harms ranking performance. Results in Table~\ref{tab:ranking_performance_verb} show that simultaneously performing both tasks does not significantly affect the model's ranking performance. \looseness=-1

\noindent{\textbf{Answer to RQ2: LLMs can not express their ranking performance in words accurately and are severely overconfident.}}
We analyze the relation between verbalized confidence predicted by models of varying sizes and actual performance for Precision@10 and $\text{NDCG-I}@10$. The results are presented in Figure~\ref{fig:ECE-Verb}. We observe that, LLMs tend to assign higher verbalized confidence across all metrics. For most samples, the verbalized confidence exceeds 0.8, indicating severe overconfidence, which aligns with previous findings on verbalized confidence in LLMs' knowledge boundary perception~\citep{xiong2023can,ni2024llms}.
As a result, the model's confidence diverges significantly from actual performance, leading to a high ECE and poor Spearman correlation. 
Compared to self-consistency-based QPP (See Table~\ref{tab:avg_results_consis}), verbalized QPP is significantly worse.
In addition, this phenomenon is not effectively mitigated as the model size increases. \looseness=-1

\definecolor{lightgray}{gray}{0.9}
\section{Teaching LLMs to Express Their Ranking Performance}
\vspace{-0.1cm}
In this section, we aim to answer \textbf{RQ3}: Can supervised training enable LLM rerankers to output calibrated ranking-quality estimates?

\subsection{Method}

\noindent\textbf{Training Objective.} 
To enable LLM reranker to perceive its own ranking quality, we construct a training set $\mathcal{T}=\{x_j, y_j\}_{j=1}^N$ where $N$ is the count of training samples. For each instance, the input $x_j$ consists of a query $q_j$ and candidate passages $D_j$. The target $y_j$ is a sequence formatted as: \texttt{<answer>} $I_j$ \texttt{</answer> <confidence>} $c_j$ \texttt{</confidence>}. 
Specifically, $I_j$ is the model's self-generated ranking list for the input $x_j$ and $c_j$ represents the model's ranking performance, calculated by evaluating $I_j$ against the ground-truth relevance labels.

We consider two forms of $c_j$:
1) \textbf{Verb-Num}: a single scalar value (e.g., ``0.85") representing a specific IR metric.
2) \textbf{Verb-List}: an ordered list of binary indicators (e.g., ``[1, 0, 1]") representing the ground-truth relevance of each passage in $I_j$, which is metric-agnostic.
Label construction overview can be seen in Figure~\ref{fig:method}. The training loss is defined as:
\begin{equation}
\mathcal{L} = \frac{1}{|\mathcal{T}|} \sum_{(x, y) \in \mathcal{T}} \left( \mathcal{L}_{\text{rank}}(x, y) + \lambda \mathcal{L}_{\text{conf}}(x, y) \right) ,
\end{equation}
where $\lambda$ is a balancing parameter used to optimize confidence expression without degrading the model’s original ranking capability.
Following~\cite{liu2025sliding}, we employ an importance-aware loss $\mathcal{L}_{\text{rank}}$ to emphasize the learning of top-ranked passages. $\mathcal{L}_{\text{conf}}$ is the standard cross-entropy loss calculated over the tokens within the \texttt{<confidence>} tags. Details can be found in~\S~\ref{app:training_details}. \looseness=-1

\noindent\textbf{Insufficient Human Annotation Augmentation.}
We leverage the widely used MS MARCO training set~\cite{bajaj2016ms} as our training data. Although large-scale, its ground-truth annotations are sparse, which can lead to inaccurate training signals. To address this issue, we use the powerful reasoning model Qwen3-32B~\citep{yang2025qwen3} to augment the relevance annotations (See Figure~\ref{fig:prompt_annotation_aug}). For efficiency, we annotate only the top-10 passages ranked by each model for the first 40k queries. Further details are provided in~\S~\ref{app:annotation_augmentation}. \looseness=-1
\subsection{Experimental Setup}
Due to space limitation, see details in~\S~\ref{app:training_details}.

\begin{table*}[h!]
\centering
\caption{QPP performance on Trec DL19-22 datasets. Bold and underline indicate the best and second-best results per ranking model. Statistical significance ($p < 0.05$) is denoted by \textbf{*} for improvement over QPP-Gen-SFT (supervised on MS MARCO) and \textbf{\dag} for improvement over QPP-Gen (3-32B). Significance for Spearman is calculated via Steiger’s Z-test, and for ECE via a paired t-test.}
\label{tab:OOD_evaluation}
\small
\setlength{\tabcolsep}{4.5pt}
\scalebox{0.8}{\begin{tabular}{ll ccccc ccccc}
\toprule
& & \multicolumn{5}{c}{Spearman $\uparrow$} & \multicolumn{5}{c}{ECE $\downarrow$} \\
\cmidrule(lr){3-7} \cmidrule(lr){8-12}
Rankers & Methods & DL19 & DL20 & DL21 & DL22 & Avg. & DL19 & DL20 & DL21 & DL22 & Avg. \\
\midrule
\multirow{15}{*}{\shortstack[l]{Qwen2.5-7B-Instruc100}} 
& \multicolumn{11}{c}{\textit{Unsupervised}} \\
& QPP-Gen               & 0.178 & 0.479 & 0.347 & 0.497 & 0.397 & 0.288 & \textbf{0.152} & 0.234 & \underline{0.176} & \underline{0.205} \\
& QPP-Gen (3-32B)            & 0.407 & 0.534 & 0.273 & \underline{0.669} & 0.494 & \underline{0.212} & 0.239 & \underline{0.233} & 0.193 & 0.217 \\
\cdashline{2-12}
& \multicolumn{11}{c}{\textit{Supervised on MS MARCO}} \\
& BERTQPP              & 0.030 & -0.022 & 0.103 & 0.004 & 0.026 & 0.564 & 0.444 & 0.524 & 0.300 & 0.437 \\
& qppBERT-PL           & 0.164 & 0.233 & 0.189 & 0.105 & 0.167 & 0.587 & 0.460 & 0.588 & 0.341 & 0.474 \\
& QPP-Gen-SFT          & 0.253 & 0.477 & 0.450 & 0.462 & 0.423 & 0.230 & 0.297 & 0.293 & 0.362 & 0.305 \\
\cdashline{2-12}
& \multicolumn{11}{c}{\textit{Supervised on Augmented MS MARCO}} \\
& BERTQPP              & 0.318 & 0.132 & 0.402 & 0.322 & 0.295 & 0.534 & 0.414 & 0.494 & 0.274 & 0.409 \\
& qppBERT-PL           & \textbf{0.499} & 0.514 & 0.253 & 0.216 & 0.350 & 0.516 & 0.423 & 0.599 & 0.377 & 0.466 \\
& QPP-Gen-SFT          & 0.258 & 0.434 & 0.223 & 0.399 & 0.339 & 0.296 & 0.401 & 0.342 & 0.508 & 0.403 \\
\cmidrule(lr){2-12} 
\rowcolor{gray!20} \cellcolor{white}& Verb-Num             & 0.385 & \textbf{0.626} & \underline{0.475}† & \textbf{0.679}* & \textbf{0.563} & 0.510 & 0.391 & 0.455 &	0.287 & 0.394 \\
\rowcolor{gray!20} \cellcolor{white}& Verb-List            & \underline{0.423} & \underline{0.605} & \textbf{0.486}† & 0.612 & \underline{0.545} & \textbf{0.166} & \underline{0.216}*† & \textbf{0.215}* & \textbf{0.163}* & \textbf{0.188} \\
\bottomrule
\end{tabular}}
\end{table*}

\noindent\textbf{Models.}
We evaluate confidence verbalization under two training settings. 
The first setting assumes access to passage-level relevance annotations, which allows us to improve both the reranker's ranking ability and its verbalized QPP ability. 
Following prior work~\citep{liu2025sliding}, we first enhance Qwen2.5-7B-Instruct with GPT-4o-generated ranking lists, and then train it to express the quality of its own rankings. 
We denote the resulting ranking-enhanced model as Qwen2.5-7B-Instruct100.
The second setting considers scenarios such as RAG, where passage-level relevance annotations may be unavailable and only an overall ranking-quality score can be obtained. 
In this case, we do not enhance the model's ranking ability, but only train it to verbalize its ranking quality. 
We refer to this original model as Qwen2.5-7B-Instruct. \looseness=-1

\noindent\textbf{Metrics.} We use normalized \textbf{DCG@10} for ranking performance, and Spearman correlation and ECE for predicted confidence, where DCG@10 is normalized by the ideal score to keep confidence values within $[0,1]$. \looseness=-1

\noindent\textbf{Datasets.} 
We train on 20k balanced queries and evaluate on 2,500 held-out MS MARCO queries in-domain and TREC DL 2019--2022 OOD.

\noindent\textbf{Baselines.} We compare a range of SOTA unsupervised and supervised QPP methods.
For unsupervised methods, we include \textbf{QPP-Gen}~\cite{meng2025query}.
For supervised methods, we consider:
\textbf{BERTQPP}~\cite{arabzadeh2021bert}: a regression-based method that predicts QPP scores from BERT representations of the query and top-ranked document.
\textbf{qppBERT-QL}~\cite{datta2022pointwise}: splits the ranked list into chunks, predicts the number of relevant items per chunk, and averages the results with weighted importance.
\textbf{QPP-Gen-SFT}~\cite{meng2025query}: fine-tunes an LLM to improve pointwise relevance judgments.



\subsection{Results and Analysis}
Table~\ref{tab:OOD_evaluation} reports the main results under the passage-level annotation setting, where the model is first enhanced for ranking and then trained for verbalized QPP. 
Table~\ref{tab:OOD_evaluation_vanilla_model} reports the results under the ranking-quality-only setting, where the model is trained to verbalize ranking quality without additional ranking enhancement. Our findings are as follows.
\noindent\textbf{1) Verb-Num and Verb-List achieve SOTA QPP performance.} As shown in Table~\ref{tab:OOD_evaluation}, the two methods we propose outperform existing approaches on most datasets, even surpassing the powerful Qwen3-32B.
Specifically, Verb-Num often achieves the best Spearman correlation, while Verb-List attains the lowest ECE. This is consistent in in-domain evaluation (See~\S~\ref{app:in-domain analysis}). 
Compared with QPP-Gen, which requires separate inference over each retrieved passage, verbalized QPP introduces minimal overhead by generating only a few additional tokens after ranking.
\begin{figure}[h]
    \centering
    \includegraphics[width=\linewidth]{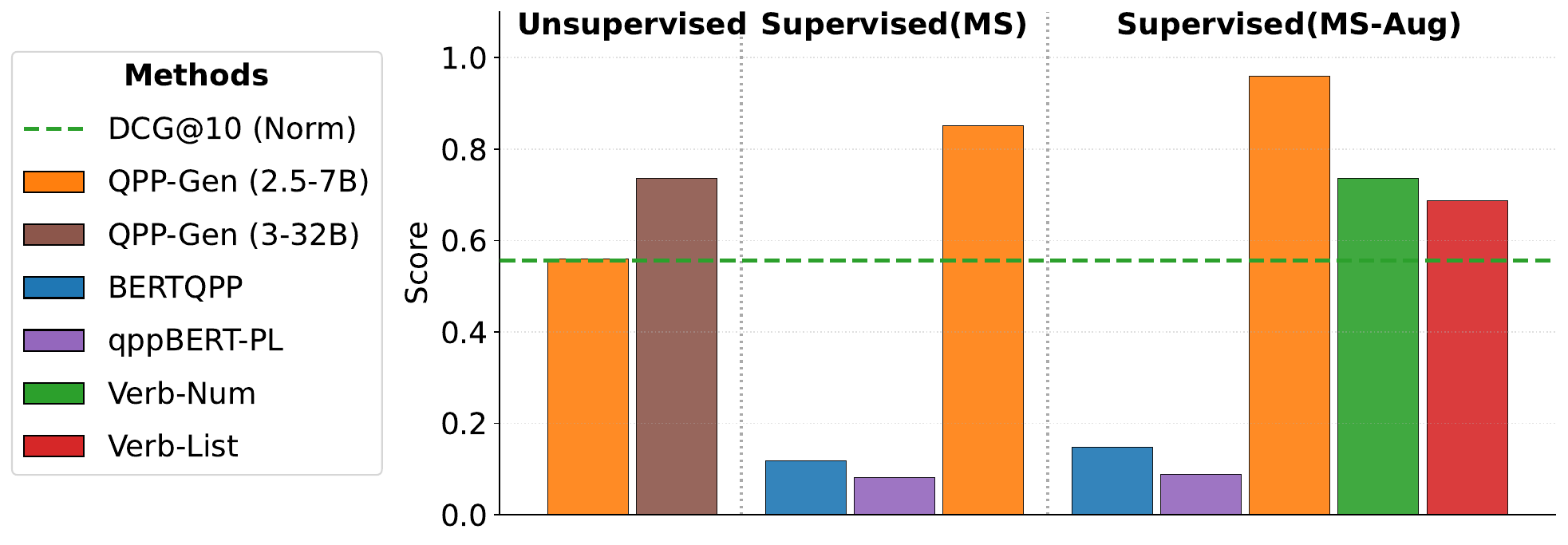}
    \caption{Average predicted scores of different methods on DL19–22 using ranking lists from Verb-Num (Qwen2.5-7B-Instruct100). The horizontal line indicates ranking performance. MS and MS-Aug denote training on the original and augmented MS MARCO annotations, respectively.}
    \label{fig:average_conf}
    \vspace{-0.1cm}
\end{figure}
\noindent\textbf{2) Compared to Verb-Num, Verb-List is better calibrated but exhibits weaker correlation.}
One possible reason is that Verb-List only needs to learn relevance judgments, without explicitly modeling complex metric computations. Consequently, its predicted confidence is numerically closer to the true performance. However, because the model treats relevance judgments at different positions equally, it fails to capture positional importance, which may lead to lower correlation. Incorporating positional weights is a promising direction for future work. Further calibration of Verb-Num could likely be addressed by using more training data.

The ranking-quality-only setting shows a similar trend, suggesting that our method can still improve verbalized QPP when only aggregate ranking-quality supervision is available. Moreover, as shown in Figure~\ref{fig:average_conf}, Qwen3-32B can produce false positives, leading to overestimate ranking performance. Details are in \S~\ref{app:in-domain analysis}. \looseness=-1

\section{Conclusion}
\vspace{-0.2cm}
In this paper, we study reranker-internal QPP, asking whether LLM rerankers can estimate the quality of their own generated rankings. We find that LLMs’ self-consistency across multiple ranking generations is competitive with the SOTA unsupervised method QPP-Gen, but they still struggle to express confidence in a human-like verbalized form. To address this, we propose two training methods, Verb-Num and Verb-List, to improve verbalized confidence.
Future directions include: (1) modeling ranking performance with multi-level relevance, (2) improving reranker-internal QPP in an unsupervised setting, and (3) exploring whether long-horizon reasoning further benefit reranker-internal QPP.\looseness=-1
\clearpage


\section*{Limitations}
First, we focus on evaluating ranking performance with relatively simple binary relevance settings, leaving reranker-internal QPP under multi-level relevance unexplored. 
Second, although Verb-Num and Verb-List improve verbalized confidence through training, we do not fully address how to improve reranker-internal QPP in an unsupervised setting. 
Third, we do not examine whether long-horizon reasoning and reflection can further benefit LLM rerankers in perceiving their own ranking performance. 
These limitations point to important directions for future research.

\section*{Ethical Considerations}
All models used in this study are open-source LLMs, and all datasets are derived from publicly available open-source resources. 
We use these models and datasets in accordance with their respective licenses and intended research purposes. 
Our experiments are conducted solely for academic research and do not involve human subjects, private user data, or sensitive personal information. 
Since our work focuses on query performance prediction and model self-awareness in ranking tasks, it does not introduce additional risks related to harmful content generation or real-world deployment. 
AI assistants were used solely for grammar or style polishing.
No original research ideas, experimental designs, analyses, or scientific claims were generated by AI tools. 
All substantive intellectual contributions and final manuscript content were produced, reviewed, and verified by the authors.




\bibliography{reference}

\appendix

\section{Related Work \label{app:related_work}} 

\subsection{Query Performance Prediction} 
Query performance prediction can be broadly categorized into pre-retrieval and post-retrieval methods. In this paper, we focus on post-retrieval methods for their effectiveness.
Existing post-retrieval QPP approaches can be broadly categorized into unsupervised and supervised methods.

\noindent\textbf{Unsupervised Methods.} A wide range of unsupervised QPP techniques have been proposed to estimate retrieval effectiveness, primarily for traditional lexical rankers such as BM25~\cite{robertson1995okapi} and query likelihood~\cite{lafferty2001document}. These methods typically exploit statistical properties of the retrieved results and can be grouped into several categories: clarity-based methods~\cite{cronen2002predicting}, which measure the divergence between query and collection language models; robustness-based methods~\cite{aslam2007query,zhou2006ranking,zhou2007query}, which assess the stability of retrieval results under query perturbations; coherence-based methods~\cite{arabzadeh2021query}, which evaluate semantic consistency among top-ranked documents; and score-based methods~\cite{shtok2012predicting,perez2010standard,cummins2011improved,tao2014query}, which derive predictive signals from the distribution of retrieval scores.
However, post-retrieval QPP methods originally developed for lexical rankers have been shown to be less effective for neural ranking models. This limitation has motivated a line of unsupervised QPP methods specifically designed for neural rankers. Representative approaches include the weighted relative information gain (WRIG) model~\cite{datta2022relative}, which estimates relative performance differences between a given query and its variants; entropy-based methods~\cite{zendel2023entropy}, which exploit the score distributions produced by neural rankers; neural-specific regularization schemes~\cite{faggioli2023towards}; coherence-based methods leveraging dense representations~\cite{vlachou2023coherence}; and pairwise preference-based methods~\cite{singh2023unsupervised}, which quantify agreement across different neural rankers.

\noindent\textbf{Supervised Methods.} In parallel, supervised QPP methods learn mappings from ranked lists to performance estimates using labeled data~\cite{zamani2018neural,hashemi2019performance,arabzadeh2021bert,datta2022deep,chen2022groupwise,datta2022pointwise,khodabakhsh2023learning}. Early neural approaches, such as NeuralQPP~\cite{zamani2018neural} and Deep-QPP~\cite{datta2022deep}, train predictive models from scratch, whereas later methods leverage pre-trained language models. For example, NQA-QPP~\cite{hashemi2019performance} and BERT-QPP~\cite{arabzadeh2021bert} fine-tune BERT~\cite{devlin2019bert}. Subsequent work, including qppBERT-PL~\cite{datta2022pointwise}, incorporates listwise document information, while BERT-groupwise-QPP~\cite{chen2022groupwise} models both cross-query and cross-document dependencies. M-QPPF~\cite{khodabakhsh2023learning} further adopts a multi-task learning framework that jointly optimizes document ranking and QPP.
More recently, Meng et al.~\cite{meng2025query} proposed QPP-Gen, which uses an LLM to judge the relevance of documents in a ranked list with respect to a query. QPP-Gen demonstrates strong performance in unsupervised settings and can be further improved through supervised training, achieving state-of-the-art results.

Most existing QPP methods treat retrieval and QPP as two separate modules. Although M-QPPF jointly models these tasks, it remains constrained within the traditional BERT-based retrieval paradigm. In contrast, this paper aims to explore a unified paradigm we term \textbf{reranker-internal QPP}, where the reranker serves as its own quality estimator.
\vspace{-0.3cm}
\subsection{LLM Knowledge Boundary Perception} 
The perception of knowledge boundaries in LLMs has been extensively studied in the question-answering (QA) domain. A reliable model should be able to recognize what it knows and what it does not, which is typically assessed by examining whether the model’s confidence in its outputs aligns with its actual performance. Existing research on confidence estimation can be broadly categorized into four approaches.

\noindent\textbf{Probabilistic Confidence}~\cite{guo2017calibration,desai2020calibration,jiang2021can,kadavath2022language,si2022prompting}. Early work measured confidence through the model’s generation probabilities. For classification tasks, Guo et al.~\cite{guo2017calibration} observed that ResNet~\cite{he2016deep} often assigns probabilities higher than the actual accuracy of its predictions and proposed temperature scaling to correct this. Subsequent studies~\cite{desai2020calibration} found that pre-trained models such as BERT~\cite{devlin2019bert} are generally well-calibrated. With the rise of generative LLMs, later research~\cite{jiang2021can,kadavath2022language,si2022prompting} revealed that models in free-form generation tasks tend to exhibit overconfidence.

\noindent\textbf{Self-consistency-based Confidence}~\cite{fomicheva2020unsupervised,Manakul2023SelfCheckGPT,kuhn2023semantic,zhang2023sac3,ding2024retrieve}. Probabilistic confidence can be influenced by generation format and is not applicable for black-box models. To address this, recent studies leverage the self-consistency of multiple model generations to estimate confidence: the more consistent the outputs across generations, the higher the model’s confidence. Early approaches~\cite{fomicheva2020unsupervised} quantified consistency via term-level overlap between outputs, while recent work~\cite{Manakul2023SelfCheckGPT,kuhn2023semantic} uses LLMs to assess semantic consistency, achieving state-of-the-art performance among unsupervised methods. Zhang et al.~\cite{zhang2023sac3} and Ding et al.~\cite{ding2024retrieve} extended this approach by measuring consistency across different query variants and languages. Wang et.al.~\cite{wang2026evaluating} bring confidence calibration into the multi-answer setting and observe that the effectiveness of self-consistency-based methods declines when dealing with multi-answer questions.

\noindent\textbf{Verbalized Confidence}~\cite{yin2023large,tian2023just,xiong2023can,ni2024llms,lin2022teaching,yang2023alignment,zhang2024r}. The capabilities of LLMs enable a new paradigm where models can express confidence in natural language. Researchers have studied whether LLMs can accurately verbalize confidence, examining both binary~\cite{yin2023large,ni2024llms} and fine-grained confidence levels~\cite{xiong2023can,tian2023just}, finding that models are highly overconfident. Parallel work has focused on training models to express confidence more accurately, demonstrating that LLMs can be taught to convey their certainty in words~\cite{lin2022teaching,yang2023alignment,zhang2024r}.

\noindent\textbf{Internal-state-based Confidence}~\cite{azaria2023internal,su2024unsupervised,chen2024inside,ni2025towards,ni2025annotation}. Beyond explicit verbalization, some studies show that a model’s internal states inherently encode information about answer correctness. Lightweight networks can extract this information~\cite{azaria2023internal,su2024unsupervised,chen2024inside}. Ni et al.~\cite{ni2025towards} found that hidden states prior to answer generation already encode confidence, and more recent work~\cite{ni2025annotation} introduces a pretraining–finetuning paradigm for confidence estimation, enabling efficient development of estimators that generalize across domains.

In this paper, we explore whether LLMs can recognize their own ranking performance which is fundamentally different compared with LLMs' self-awareness in answer factuality.


\begin{table}[t] 
\centering
\caption{Average ranking performance and self-consistency results across TREC DL19-22 datasets. The averages are weighted by the number of queries in each dataset.}
\label{tab:ranking performance}
\small 
\scalebox{0.77}{\begin{tabular}{ll cc}
\toprule
\textbf{Metrics} & \textbf{Models} & \textbf{Ranking Scores} & \textbf{Consistency} \\
\midrule
\multirow{5}{*}{$\text{NDCG-I}$@10} 
& BM25        & 0.374 & - \\
& Llama3.1-8B & 0.493 & 0.642 \\
& Qwen2.5-7B  & 0.496 & 0.641 \\
& Qwen2.5-14B & 0.515 & 0.664 \\
& Qwen2.5-32B & \textbf{0.555} & \textbf{0.700} \\
\midrule
\multirow{5}{*}{Precision@10} 
& BM25        & 0.293 & - \\
& Llama3.1-8B & 0.346 & 0.508 \\
& Qwen2.5-7B  & 0.349 & 0.511 \\
& Qwen2.5-14B & 0.356 & 0.500 \\
& Qwen2.5-32B & \textbf{0.385} & \textbf{0.537} \\
\midrule
\multirow{5}{*}{MRR@10} 
& BM25        & 0.511 & - \\
& Llama3.1-8B & 0.703 & 0.877 \\
& Qwen2.5-7B  & 0.710 & 0.868 \\
& Qwen2.5-14B & 0.707 & 0.898 \\
& Qwen2.5-32B & \textbf{0.738} & \textbf{0.903} \\
\bottomrule
\end{tabular}}
\end{table}

\section{Datasets \label{app:datasets}}
TREC DL 19-22 are built on the MS MARCO passage collections: TREC-DL 2019–2020 uses MS MARCO V1, while 2021–2022 uses MS MARCO V2, containing approximately 8.8M and 138M passages, respectively.
Each query–passage pair is annotated with one of four relevance levels: perfectly relevant (3), highly relevant (2), related (1), or irrelevant (0). Following common practice in TREC-DL studies~\cite{meng2025query,arabzadeh2021bert}, we binarize relevance labels by treating passages with relevance $\ge$ 2 as positive. The statistics of datasets can be seen in Table~\ref{tab:dataset_stats}.
\begin{table*}[ht]
\centering
\caption{Statistics for the TREC 2019–2022 Deep Learning (DL) tracks. Rel. labels represent the average number of passages at each relevance level (0/1/2/3) for each query. Top 100 Rel. labels denote the number of passages at each relevance level within the top-100 passages retrieved by BM25.}
\label{tab:dataset_stats}
\begin{tabular}{lcrr}
\toprule
Track & \# Queries & Rel. labels & Top 100 Rel. labels \\
\midrule
DL19 & 43 & 120 / 37 / 42 / 16 & 68 / 12 / 13 / 7 \\
DL20 & 54 & 144 / 36 / 19 / 12 & 76 / 12 / 6 / 6 \\
DL21 & 53 & 82 / 58 / 44 / 20 & 66 / 18 / 11 / 6 \\
DL22 & 76 & 3769 / 687 / 606 / 22 & 76 / 17 / 6 / 2 \\
\bottomrule
\end{tabular}
\end{table*}

\begin{table*}[t]
\centering
\caption{QPP performance of self-consistency and QPP-Gen. ``Avg." reports the weighted average over Trec DL19-22, where weights are proportional to the number of queries in each dataset. ``–" indicates that all predicted values are 1, so the correlation coefficient cannot be computed.}
\label{tab:spearman_and_ece}
\resizebox{0.95\textwidth}{!}{%
\begin{tabular}{lllrrrrrrrrrr}
\toprule
\multirow{2}{*}{Metrics} & \multirow{2}{*}{Models} & \multirow{2}{*}{Methods} & \multicolumn{5}{c}{Spearman $\uparrow$} & \multicolumn{5}{c}{ECE $\downarrow$} \\
\cmidrule(lr){4-8} \cmidrule(lr){9-13}
 &  &  & DL19 & DL20 & DL21 & DL22 & Avg. & DL19 & DL20 & DL21 & DL22 & Avg. \\
\midrule
\multirow{8}{*}{$\text{NDCG-I}$@10} & \multirow{2}{*}{Llama3.1-8B-Instruct} & QPP-Gen & 0.248 & \textbf{0.235} & 0.381 & 0.224 & 0.268 & 0.331 & 0.357 & 0.338 & 0.480 & 0.389 \\
 &  & Self-Consis & \textbf{0.471} & 0.234 & \textbf{0.419} & \textbf{0.277} & \textbf{0.337} & \textbf{0.149} & \textbf{0.184} & \textbf{0.103} & \textbf{0.245} & \textbf{0.179} \\
\cmidrule(lr){2-13}
 & \multirow{2}{*}{Qwen2.5-7B-Instruct} & QPP-Gen & 0.291 & \textbf{0.355} & 0.210 & 0.297 & 0.289 & 0.251 & 0.176 & 0.208 & \textbf{0.183} & 0.200 \\
 &  & Self-Consis & \textbf{0.350} & 0.247 & \textbf{0.429} & \textbf{0.326} & \textbf{0.336} & \textbf{0.154} & \textbf{0.140} & \textbf{0.112} & 0.244 & \textbf{0.171} \\
\cmidrule(lr){2-13}
 & \multirow{2}{*}{Qwen2.5-14B-Instruct} & QPP-Gen & \textbf{0.470} & 0.377 & \textbf{0.300} & \textbf{0.361} & 0.371 & 0.191 & 0.142 & 0.223 & 0.255 & 0.208 \\
 &  & Self-Consis & 0.364 & \textbf{0.541} & 0.253 & 0.343 & \textbf{0.373} & \textbf{0.119} & \textbf{0.088} & \textbf{0.142} & \textbf{0.240} & \textbf{0.158} \\
\cmidrule(lr){2-13}
 & \multirow{2}{*}{Qwen2.5-32B-Instruct} & QPP-Gen & \textbf{0.371} & 0.214 & 0.369 & 0.258 & 0.295 & 0.178 & 0.162 & 0.215 & 0.266 & 0.212 \\
 &  & Self-Consis & 0.355 & \textbf{0.510} & \textbf{0.400} & \textbf{0.328} & \textbf{0.394} & \textbf{0.088} & \textbf{0.096} & \textbf{0.133} & \textbf{0.236} & \textbf{0.150} \\
\midrule
\multirow{8}{*}{Precision@10} & \multirow{2}{*}{Llama3.1-8B-Instruct} & QPP-Gen & 0.355 & \textbf{0.487} & \textbf{0.495} & \textbf{0.403} & \textbf{0.436} & 0.428 & 0.487 & 0.442 & 0.579 & 0.496 \\
 &  & Self-Consis & \textbf{0.591} & 0.268 & 0.340 & 0.364 & 0.379 & \textbf{0.108} & \textbf{0.148} & \textbf{0.091} & \textbf{0.263} & \textbf{0.166} \\
\cmidrule(lr){2-13}
 & \multirow{2}{*}{Qwen2.5-7B-Instruct} & QPP-Gen & 0.275 & \textbf{0.531} & 0.325 & \textbf{0.438} & 0.403 & 0.240 & \textbf{0.091} & 0.206 & \textbf{0.113} & \textbf{0.154} \\
 &  & Self-Consis & \textbf{0.558} & 0.294 & \textbf{0.448} & 0.425 & \textbf{0.424} & \textbf{0.117} & 0.132 & \textbf{0.096} & 0.267 & 0.166 \\
\cmidrule(lr){2-13}
 & \multirow{2}{*}{Qwen2.5-14B-Instruct} & QPP-Gen & \textbf{0.624} & 0.642 & \textbf{0.361} & \textbf{0.569} & \textbf{0.548} & 0.200 & 0.200 & 0.292 & 0.282 & 0.249 \\
 &  & Self-Consis & 0.415 & \textbf{0.702} & 0.152 & 0.418 & 0.423 & \textbf{0.103} & \textbf{0.108} & \textbf{0.139} & \textbf{0.238} & \textbf{0.158} \\
\cmidrule(lr){2-13}
 & \multirow{2}{*}{Qwen2.5-32B-Instruct} & QPP-Gen & \textbf{0.552} & 0.591 & \textbf{0.453} & \textbf{0.571} & \textbf{0.544} & 0.195 & 0.230 & 0.308 & 0.257 & 0.251 \\
 &  & Self-Consis & 0.358 & \textbf{0.624} & 0.342 & 0.467 & 0.454 & \textbf{0.118} & \textbf{0.171} & \textbf{0.123} & \textbf{0.235} & \textbf{0.171} \\
\midrule
\multirow{8}{*}{MRR@10} & \multirow{2}{*}{Llama3.1-8B-Instruct} & QPP-Gen & -0.073 & 0.263 & \textbf{0.243} & \textbf{0.220} & 0.180 & 0.142 & 0.207 & 0.255 & 0.434 & 0.282 \\
 &  & Self-Consis & \textbf{0.166} & \textbf{0.270} & 0.241 & 0.216 & \textbf{0.225} & \textbf{0.126} & \textbf{0.188} & \textbf{0.205} & \textbf{0.309} & \textbf{0.221} \\
\cmidrule(lr){2-13}
 & \multirow{2}{*}{Qwen2.5-7B-Instruct} & QPP-Gen & \textbf{0.180} & 0.243 & \textbf{0.263} & \textbf{0.316} & \textbf{0.260} & 0.240 & 0.242 & 0.250 & \textbf{0.293} & 0.261 \\
 &  & Self-Consis & 0.037 & \textbf{0.366} & 0.139 & 0.229 & 0.204 & \textbf{0.131} & \textbf{0.160} & \textbf{0.218} & 0.310 & \textbf{0.219} \\
\cmidrule(lr){2-13}
 & \multirow{2}{*}{Qwen2.5-14B-Instruct} & QPP-Gen & - & \textbf{0.364} & 0.355 & \textbf{0.456} & - & 0.158 & 0.174 & 0.159 & \textbf{0.316} & 0.215 \\
 &  & Self-Consis & 0.314 & 0.255 & \textbf{0.490} & 0.213 & 0.307 & \textbf{0.086} & \textbf{0.139} & \textbf{0.121} & 0.370 & \textbf{0.202} \\
\cmidrule(lr){2-13}
 & \multirow{2}{*}{Qwen2.5-32B-Instruct} & QPP-Gen & -0.068 & \textbf{0.303} & 0.363 & \textbf{0.342} & \textbf{0.260} & 0.124 & \textbf{0.192} & 0.189 & 0.344 & 0.229 \\
 &  & Self-Consis & \textbf{0.331} & 0.045 & \textbf{0.411} & 0.119 & 0.210 & \textbf{0.110} & 0.211 & \textbf{0.142} & \textbf{0.276} & \textbf{0.197} \\
\bottomrule
\end{tabular}%
}
\end{table*}

\begin{table}[htbp]
\centering
\caption{Ranking performance on TREC DL datasets. The values represent the weighted mean across DL19-22 datasets. Full refers to ranking only. Full+Verb refers to ranking while also outputting confidence.}
\label{tab:ranking_performance_verb}
\resizebox{\columnwidth}{!}{ 
\begin{tabular}{llccc}
\toprule
\textbf{Metric} & \textbf{Types} & \textbf{Qwen-7B} & \textbf{Qwen-14B} & \textbf{Qwen-32B} \\
\midrule
\multirow{2}{*}{$\text{NDCG-I}$@10} 
& Full & 0.496 & 0.515 & \textbf{0.555} \\
& Full+Verb & \textbf{0.509} & \textbf{0.516} & 0.542 \\ 
\bottomrule
\end{tabular}
}
\end{table}

\section{Training Details. \label{app:training_details}}
\noindent\textbf{Ranking Loss.} Following~\cite{liu2025sliding}, we employ an importance-aware loss $\mathcal{L}_{\text{rank}}$ to emphasize the learning of top-ranked passages:
\begin{equation}
\mathcal{L}_{\text{rank}} = -\sum_{i \in \mathcal{Y}_{ans}} w_i \log P_{\theta}(y_i \mid x, y_{<i}) \, ,
\label{eq:ranking_loss}
\end{equation}
where $\mathcal{Y}_{ans}$ denotes the token indices within the \texttt{<answer>} tags. The weight $w_i$ is assigned as:
\begin{equation}
w_i = 
\begin{cases} 
1 + \frac{1}{\log_2(p_i+1)}, & y_i \in \text{passage IDs}, \\
1, & \text{otherwise},
\end{cases}
\end{equation}
where $p_i$ represents the rank of the passage corresponding to token $y_i$ in $I_j$. For non-passage tokens, the weight is set to $1$.

\noindent\textit{Confidence Loss.} $\mathcal{L}_{\text{conf}}$ is the standard cross-entropy loss calculated over the tokens within the \texttt{<confidence>} tags ($\mathcal{Y}_{conf}$):
\begin{equation}
\mathcal{L}_{\text{conf}} = -\sum_{i \in \mathcal{Y}_{conf}} \log P_{\theta}(y_i \mid x, y_{<i}) \, .
\end{equation}

\begin{table*}[h]
\centering
\caption{The average number of relevant documents among the top-10 passages in the ranking lists annotated by humans and by Qwen3 on MS MARCO-Train. ``Used" denotes the training data actually used after label balancing.}
\label{tab:annotaiton_train_data}
\small 
\scalebox{0.95}{\begin{tabular}{lcccc}
\toprule
\multirow{2}{*}{\textbf{Annotators}} & \multicolumn{2}{c}{Qwen2.5-7B-Instruct} & \multicolumn{2}{c}{Qwen2.5-7B-Instruct100} \\
\cmidrule(lr){2-3} \cmidrule(lr){4-5} 
 & \# Queries & \# Rel. Passages & \# Queries & \# Rel. Passages \\ \midrule
Human      & 40,000 & 0.387 & 40,000 & 0.512 \\ 
Qwen3      & 40,000 & 5.244 & 40,000 & 7.002 \\
Qwen3+Hum  & 40,000 & 5.273 & 40,000 & 7.043 \\
\midrule
Used       & 20,000 & 4.586 & 20,000 & 5.051 \\
\bottomrule
\end{tabular}}
\label{tab:annotation_res}
\end{table*}

\noindent\textbf{Detailed Training Settings.} For listwise methods—Verb-Num, Verb-List, BERTQPP, and qppbert-PL—training effectiveness is ensured by performing label balancing across DCG@10 ranges, resulting in 20k queries for training (see the \textit{Used} row of Table~\ref{tab:annotaiton_train_data}). The confidence loss weight is set to $\lambda = 100$ for Verb-Num and $\lambda = 10$ for Verb-List. Training is conducted using the AdamW optimizer with a batch size of 32 and an initial learning rate of $5 \times 10^{-6}$.
For QPP-Gen-SFT, we follow~\cite{meng2025query} and adopt LoRA with $r = 8$ and $\alpha = 16$. For all trainable baselines, we use training labels under two settings: the original MS MARCO-Train annotations and the augmented annotations. Each method is trained for 5 epochs, and the checkpoint with the best Spearman coefficient is selected for evaluation. \looseness=-1

\begin{table*}[h]
\centering
\caption{Performance of Verb-Num and Verb-List in the in-domain evaluation using MS MARCO.}
\label{tab:in_domain_evaluation}
\small 
\begin{tabular}{lcccc}
\toprule
\multirow{2}{*}{\textbf{Metrics}} & \multicolumn{2}{c}{Qwen2.5-7B-Instruct} & \multicolumn{2}{c}{Qwen2.5-7B-Instruct100} \\
\cmidrule(lr){2-3} \cmidrule(lr){4-5} 
 & Verb-Num & Verb-List & Verb-Num & Verb-List \\ \midrule
DCG@10(Norm) & 0.470 & 0.462 & 0.551 & 0.540 \\ 
Confidence  & 0.570 & 0.554 & 0.634 & 0.559 \\
Spearman    & 0.823 & 0.769 & 0.766 & 0.756 \\
ECE         & 0.102 & 0.089 & 0.087 & 0.064 \\
\bottomrule
\end{tabular}
\end{table*}

\noindent\textbf{Evaluation Datasets.}
\textit{In-domain.} We use Qwen3-32B to supplement relevance annotations for additional 5,000 MS MARCO queries. To evaluate QPP across different query difficulties, we perform label balancing based on DCG@10 and employ 2,500 queries for in-domain evaluation.
\textit{Out-of-domain (OOD).} For OOD evaluation, we use TREC DL19-22. The distributions of queries with respect to DCG@10 in these datasets are shown in Figure~\ref{fig:DCG@10_distribution}.
For in-domain evaluation, ranking performance is computed based on the enhanced ground-truth annotations. For OOD evaluation, it is computed directly using the TREC ground-truth annotations.
Since Verb-Num and Verb-List can only perform QPP on their own generated ranking lists, it is not possible to make a perfectly controlled comparison between the two methods. In this paper, we directly compare their QPP performance on their respective ranking lists. As the overall ranking performance of Verb-Num and Verb-List is very similar, this comparison is acceptable. For all other methods, QPP is conducted on the ranking list generated by Verb-Num.

\subsection{Annotation Augmentation \label{app:annotation_augmentation}}
We leverage Qwen3-32B for relevance annotation because of its strong reasoning capabilities, which enable relatively accurate relevance judgments. At the same time, we avoid using larger models due to computational cost considerations.
The annotation results are shown in Table~\ref{tab:annotation_res}. For Qwen2.5-7B-Instruct, the model judged an average of 5.2 passages among the top-10 ranked results as relevant, whereas for Qwen2.5-7B-Instruct100, the average number of relevant passages increased to 7.
However, Qwen3-32B can misclassify irrelevant documents as relevant, leading to an inflated number of positive labels, as illustrated in Figure~\ref{fig:DCG@10_distribution}. Training on such data causes the model to overestimate ranking performance. To mitigate this issue, we partition all samples into 10 intervals according to normalized DCG@10 values ranging from 0 to 1, and perform sample balancing across these intervals. After balancing, we retain 20,000 samples for training. A comparison of the data distributions before and after balancing is shown in Figure~\ref{fig:DCG@10_distribution}.

\begin{table*}[t]
\centering
\caption{QPP performance on Trec DL19-22 datasets. Bold and underline indicate the best and second-best results per ranking model. Statistical significance ($p < 0.05$) is denoted by \textbf{*} for improvement over QPP-Gen-SFT (supervised on MS MARCO) and \textbf{\dag} for improvement over QPP-Gen (3-32B). Significance for Spearman is calculated via Steiger’s Z-test, and for ECE via a paired t-test.}
\label{tab:OOD_evaluation_vanilla_model}
\small
\setlength{\tabcolsep}{4.5pt}
\scalebox{0.8}{\begin{tabular}{ll ccccc ccccc}
\toprule
& & \multicolumn{5}{c}{Spearman $\uparrow$} & \multicolumn{5}{c}{ECE $\downarrow$} \\
\cmidrule(lr){3-7} \cmidrule(lr){8-12}
Rankers & Methods & DL19 & DL20 & DL21 & DL22 & Avg. & DL19 & DL20 & DL21 & DL22 & Avg. \\
\midrule
\multirow{15}{*}{Qwen2.5-7B-Instruct} 
& \multicolumn{11}{c}{\textit{Unsupervised}} \\
& QPP-Gen              & 0.333 & 0.412 & 0.291 & 0.430 & 0.375 & 0.241 & \textbf{0.151} & 0.234 & 0.164 & 0.192 \\
& QPP-Gen (3-32B)            & \underline{0.611} & 0.608 & \underline{0.365} & 0.605 & 0.551 & \underline{0.161} & \underline{0.156} & \underline{0.225} & 0.187 & \underline{0.184} \\
\cdashline{2-12}
& \multicolumn{11}{c}{\textit{Supervised on MS MARCO}} \\
& BERTQPP              & 0.135 & 0.252 & 0.270 & -0.060 & 0.129 & 0.419 & 0.345 & 0.369 & 0.133 & 0.293 \\
& qppBERT-PL          & -0.085 & 0.254 & -0.077 & 0.229 & 0.103 & 0.452 & 0.384 & 0.305 & \textbf{0.077} & 0.275 \\
& QPP-Gen-SFT          & 0.496 & 0.567 & 0.354 & 0.509 & 0.484 & 0.236 & 0.260 & 0.334 & 0.370 & 0.310 \\
\cdashline{2-12}
& \multicolumn{11}{c}{\textit{Supervised on Augmented MS MARCO}} \\
& BERTQPP              & 0.414 & 0.272 & 0.310 & 0.432 & 0.362 & 0.384 & 0.312 & 0.336 & \underline{0.105} & 0.262 \\
& qppBERT-PL           & 0.521 & 0.650 & -0.024 & 0.196 & 0.315 & 0.406 & 0.339 & 0.419 & 0.177 & 0.316 \\
& QPP-Gen-SFT          & 0.478 & 0.410 & 0.361 & 0.288 & 0.370 & 0.338 & 0.390 & 0.437 & 0.577 & 0.454 \\
\cmidrule(lr){2-12} 
\rowcolor{gray!20} \cellcolor{white} & Verb-Num             & \textbf{0.636} & \textbf{0.713}*† & \textbf{0.414} & \underline{0.617} & \textbf{0.596} & 0.385 & 0.323 & 0.328 & 0.151\textbf{*} & 0.278 \\
\rowcolor{gray!20} \cellcolor{white} & Verb-List           & 0.484 & \underline{0.699}* & 0.335 & \textbf{0.677}*† & \underline{0.565} & \textbf{0.141}* & 0.175* & \textbf{0.202}* & 0.187* & \textbf{0.179} \\
\bottomrule
\end{tabular}}
\end{table*}

\subsection{Detailed Analysis \label{app:in-domain analysis}}
\begin{figure}[h]
    \centering
    \includegraphics[width=\linewidth]{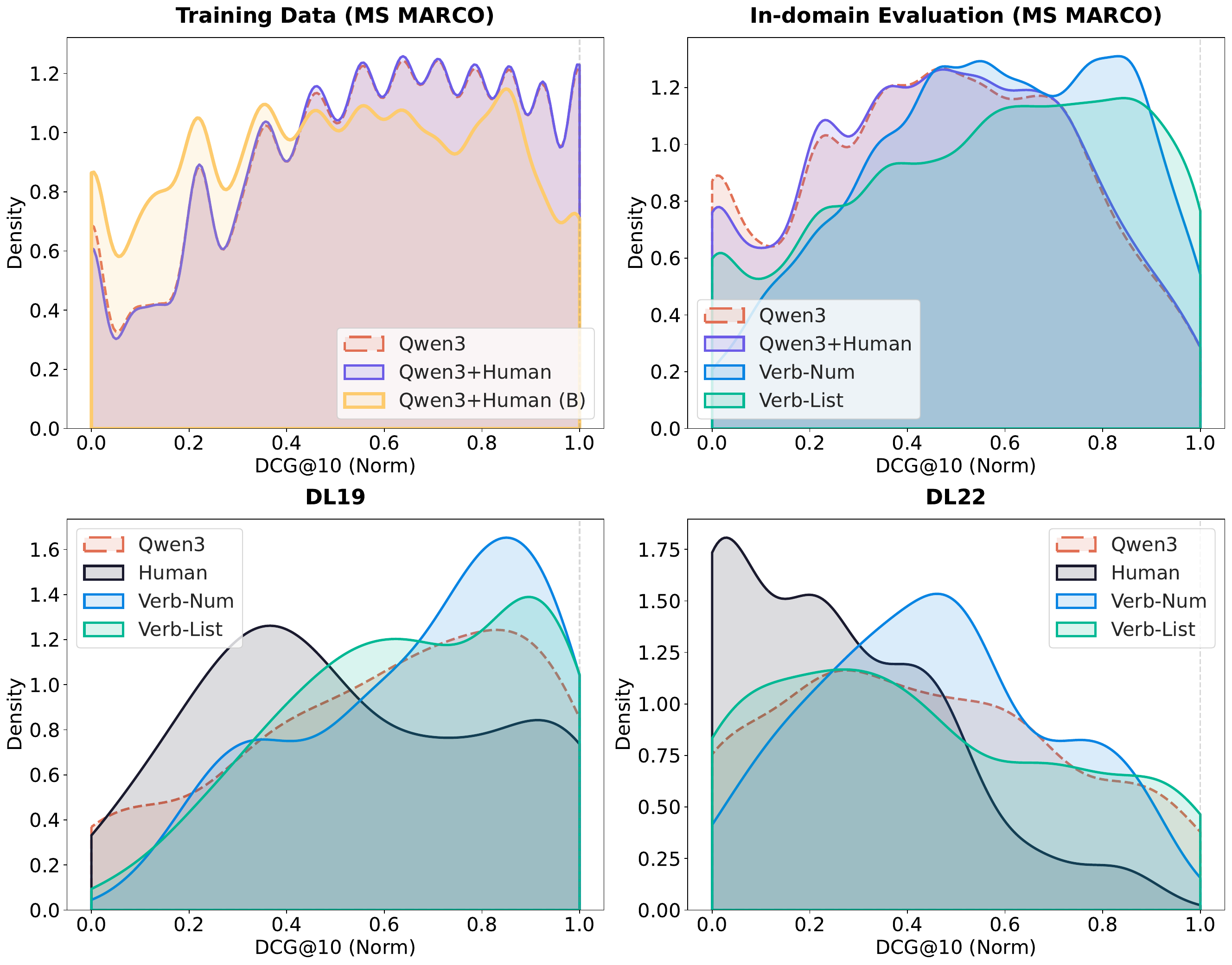}
    \caption{The query distributions corresponding to the predicted DCG@10 scores and the ground-truth scores for different methods on the in-domain and OOD datasets. All results are based on Qwen2.5-7B-Instruct. (B) represents the actual distribution of the training data used after label balancing. }
    \label{fig:DCG@10_distribution}
\end{figure}

\textbf{In in-domain evaluation, the QPP performance of both Verb-Num and Verb-List is very strong.} As shown in Table~\ref{tab:in_domain_evaluation}, the Spearman correlation between the predicted scores and true ranking performance exceeds 0.75 for both methods, with the ECE consistently below 0.1. This suggests that the models effectively perceive their own ranking performance. When comparing the two methods, Verb-Num achieves a higher Spearman correlation than Verb-List, but its ECE is slightly worse. Unlike Verb-List, predicting the absolute ranking performance in Verb-Num requires the model to compute DCG@10, which may require more training data. In contrast, Verb-List only needs to perform relevance judgments without calculating the exact score. As seen in the first row, second column of Figure~\ref{fig:DCG@10_distribution}, Verb-List closely matches the true score distribution. However, since Verb-List assigns equal weight to all top-10 passages and does not prioritize higher-ranked passages, its Spearman correlation is lower.

As shown in Table~\ref{tab:in_domain_evaluation}, both methods still exhibit slight overconfidence in the in-domain setting. However, the overall ECE remains very small. In the OOD setting (See Table~\ref{tab:OOD_evaluation}), both the ECE and Spearman correlation become weaker, although the results are still state-of-the-art (SOTA). To better understand why the models remain slightly overconfident after training, we further analyze the annotation results and model predictions, as illustrated in Figure~\ref{fig:DCG@10_distribution}. We observe that, on TREC, Qwen3-32B generally overestimates the ranking performance compared with human annotations, which in turn leads to the overconfidence observed in Verb-Num and Verb-List. In addition, Verb-List is overall more consistent with Qwen3’s judgments, which also explains why its ECE is lower than that of Verb-Num.
This issue does not stem from our method itself, but rather from limitations in annotation resources and cost. We believe that with perfectly accurate annotations and sufficient training data, both Verb-Num and Verb-List could achieve better performance. Addressing this limitation represents a promising direction for future work.

\begin{figure*}[h]
    \centering
    \includegraphics[width=0.9\linewidth]{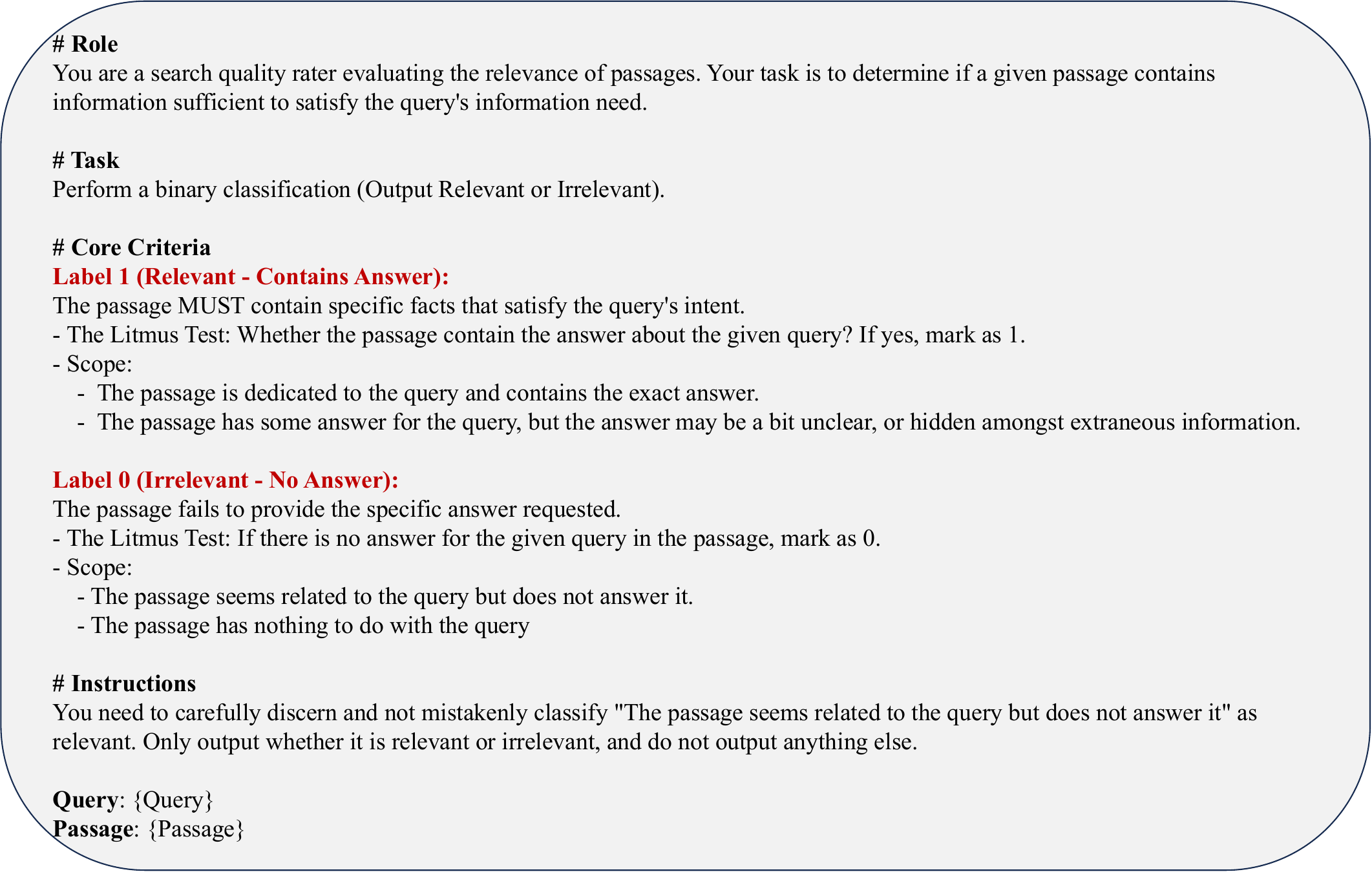}
    \caption{Instruction for Qwen3-32B Annotation Augmentation.}
    \label{fig:prompt_annotation_aug}
\end{figure*}

\begin{figure*}
    \centering
    \includegraphics[width=0.9\linewidth]{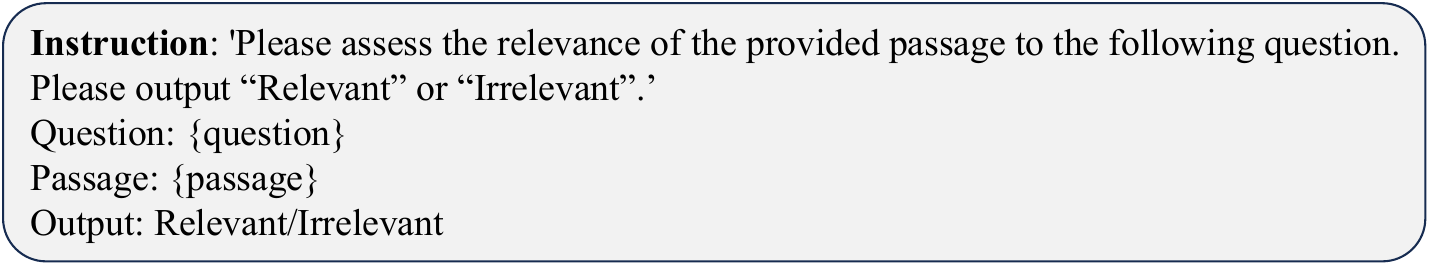}
    \caption{Instruction for QPP-Gen.}
    \label{fig:prompt_qpp_gen}
\end{figure*}

\begin{figure*}
    \centering
    \includegraphics[width=0.9\linewidth]{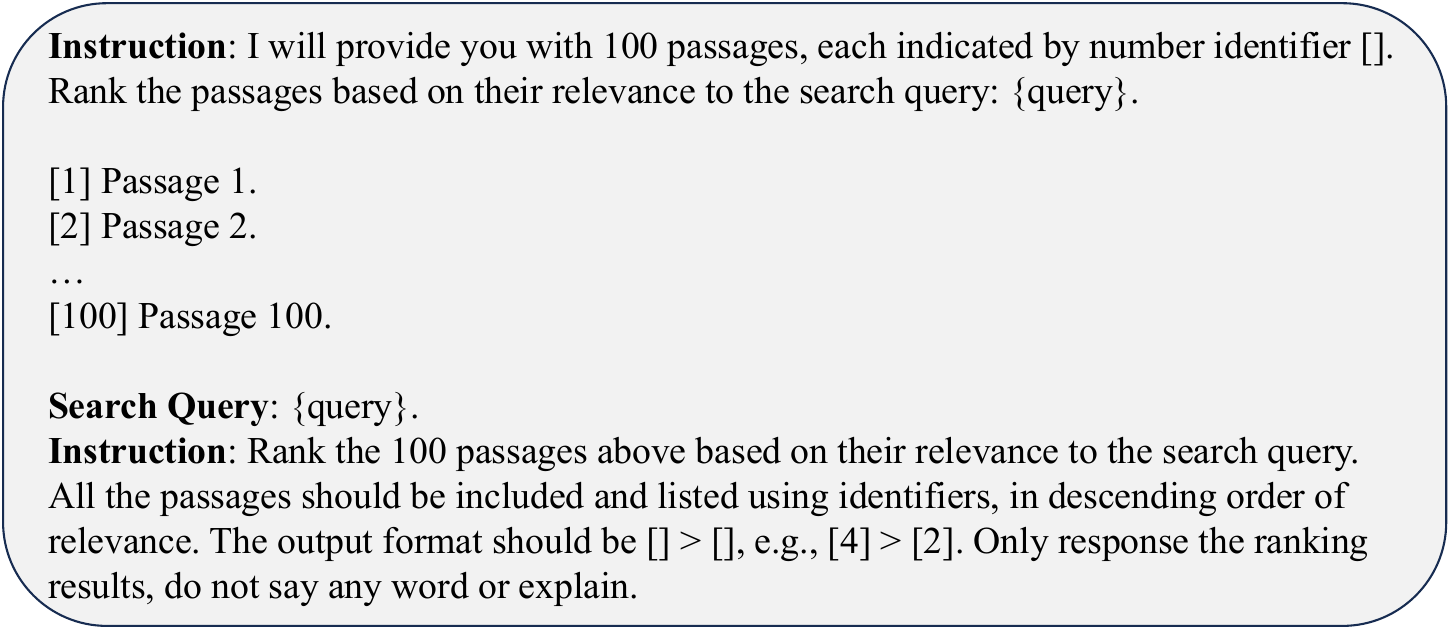}
    \caption{Instruction for sequence-to-sequence ranking.}
    \label{fig:prompt_ranking}
\end{figure*}

\begin{figure*}
    \centering
    \includegraphics[width=0.9\linewidth]{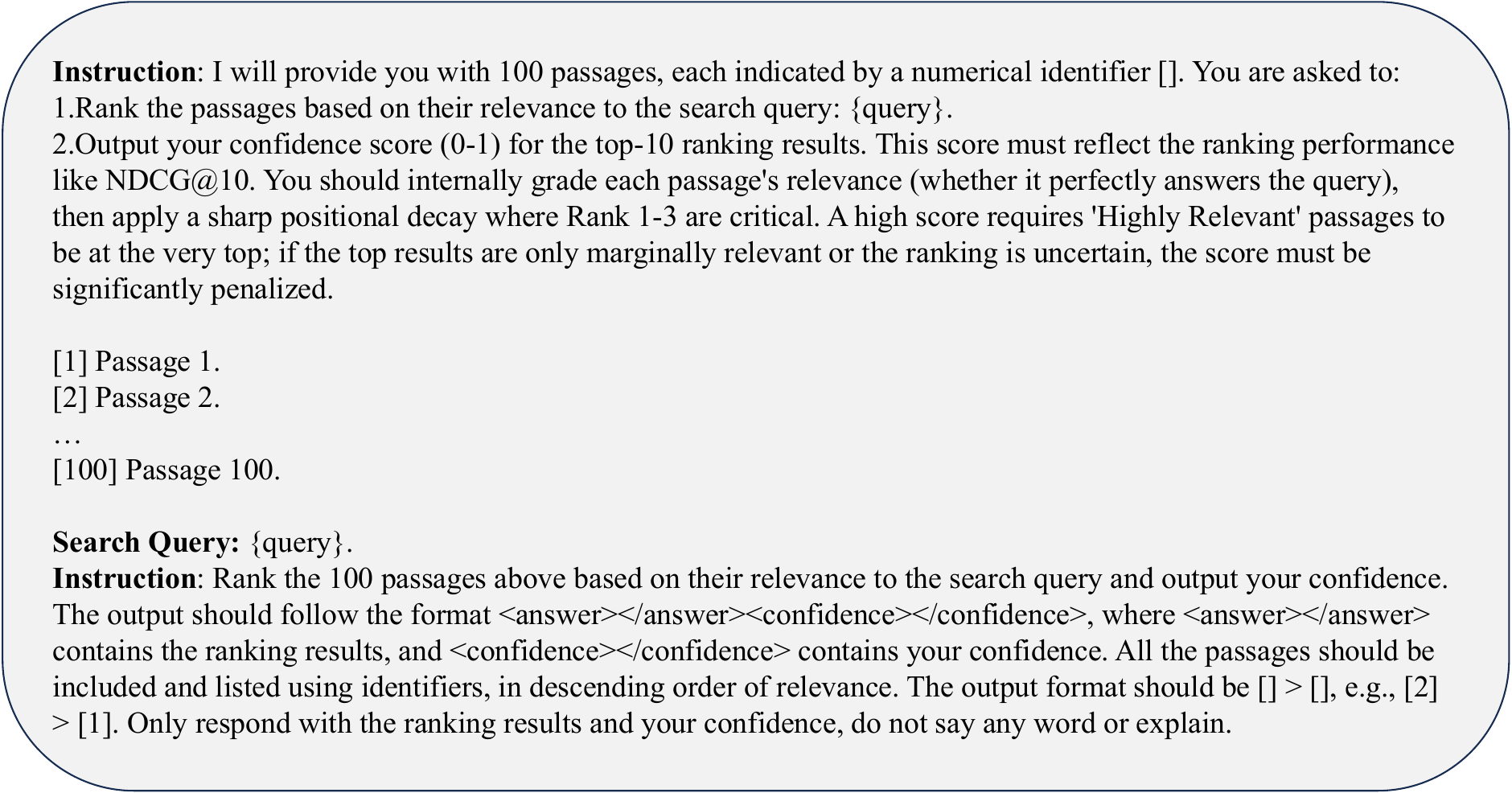}
    \caption{Instruction for Verb-Num.}
    \label{fig:prompt_verb_num}
\end{figure*}

\begin{figure*}
    \centering
    \includegraphics[width=0.9\linewidth]{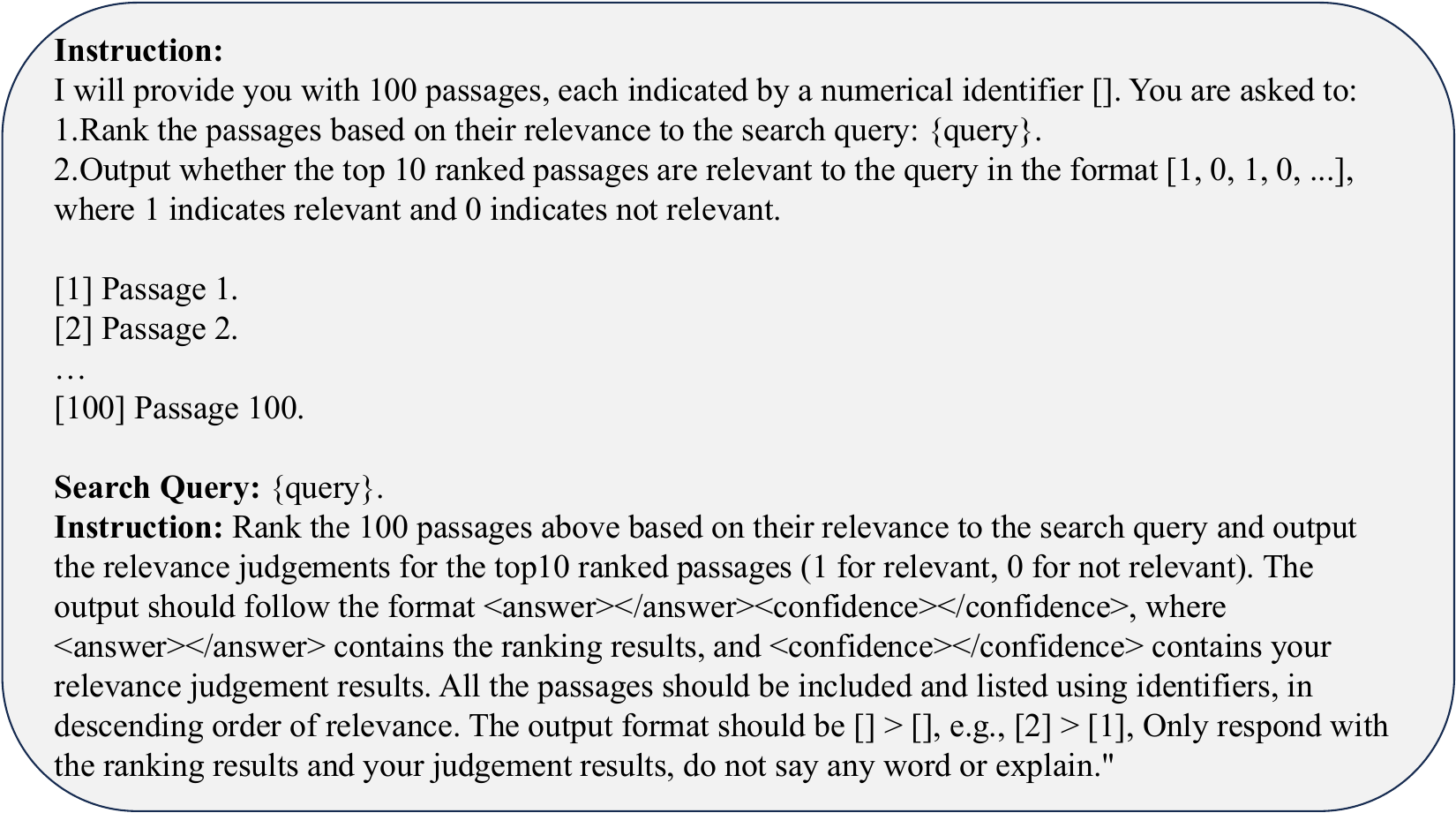}
    \caption{Instruction for Verb-List.}
    \label{fig:prompt_verb_list}
\end{figure*}

\end{document}